\documentclass[a4paper,11pt]{article}

\usepackage{graphicx}
\usepackage{times}

\usepackage{amsmath,amssymb}
\usepackage{caption}
\usepackage{url}
\usepackage[superscript]{cite}
\usepackage[a4paper, total={160mm,247mm}]{geometry}

\usepackage{titlesec}
\renewcommand\thesection{\arabic{section}.~}
\titleformat{\subsubsection}[runin]
{\normalfont\bfseries}{\thesubsubsection}{1em}{}
\begin{document}

\begin{center}
{

\textbf{\Large Signature of transition to supershear rupture speed in coseismic off-fault damage zone} \\[20pt]

Jorge Jara$^{1,*}$, Lucile Bruhat$^{1}$, Marion Y. Thomas$^{2}$, Sol\`ene L. Antoine$^{3}$, Kurama Okubo$^{4}$, Esteban Rougier$^{5}$, Ares J. Rosakis$^{6}$, Charles G. Sammis$^{7}$, Yann Klinger$^{3}$, Romain Jolivet$^{1,8}$ and Harsha S. Bhat$^{1}$ \\[20pt]

\bf To appear in \textit{Proceedings of the Royal Society A: Mathematical, Physical and Engineering Sciences}}
\end{center}
\vspace{0.2cm}

\begin{enumerate}
\footnotesize 
\itemsep 0em
\item Laboratoire de G\'eologie, D\'epartement de G\'eosciences, \'Ecole Normale Sup\'erieure, CNRS, UMR 8538, PSL Universit\'e, Paris, France
\item Institut des Sciences de la Terre de Paris, Sorbonne Universit\'e, CNRS, UMR 7193, Paris, France
\item Universit\'e de Paris, Institut de Physique du Globe de Paris, CNRS, 75005 Paris, France
\item National Research Institute for Earth Science and Disaster Resilience, 3-1 Tennnodai, Tsukuba 305-0006, Ibaraki, Japan
\item EES-17 – Earth and Environmental Sciences Division, Los Alamos National Laboratory, Los Alamos, NM, USA
\item Graduate Aerospace Laboratories, California Institute of Technology, Pasadena, California, 91125, USA
\item Department of Earth Sciences, University of Southern California, Los Angeles, CA 90089, USA
\item Institut Universitaire de France, 1 rue Descartes, 75005 Paris, France
\item[*] \textbf{Corresponding author}: \texttt{jara@geologie.ens.fr}
\end{enumerate}

\noindent

\noindent{\bf Most earthquake ruptures propagate at speeds below the shear wave velocity within the crust, but in some rare cases, ruptures reach supershear speeds. The physics underlying the transition of natural subshear earthquakes to supershear ones is currently not fully understood. Most observational studies of supershear earthquakes have focused on determining which fault segments sustain fully-grown supershear ruptures. Experimentally cross-validated numerical models have identified some of the key ingredients required to trigger a transition to supershear speed. However, the conditions for such a transition in nature are still unclear, including the precise location of this transition. In this work, we provide theoretical and numerical insights to identify the precise location of such a transition in nature. We use fracture mechanics arguments with multiple numerical models to identify the signature of supershear transition in coseismic off-fault damage. We then cross-validate this signature with high-resolution observations of fault zone width and early aftershock distributions. We confirm that the location of the transition from subshear to supershear speed is characterized by a decrease in the width of the coseismic off-fault damage zone. We thus help refine the precise location of such a transition for natural supershear earthquakes.
}

\vspace{0.2cm}
\noindent

\section{Introduction} 

While most earthquake ruptures propagate at speeds below the shear wave speed (sub-Rayleigh regime), some earthquakes can occasionally accelerate above the shear wave speed. Such events are known as supershear earthquakes. The earthquake rupture speed, and specifically its abrupt changes, control the high-frequency radiation \cite{Madariaga83, Spudich84_2}. As an earthquake goes to supershear speeds, it manifests Mach fronts that produce unusually large ground motion at distances far from the fault \cite{Bernard05, Dunham05, Dunham08}. Rupture speed also governs the spatial extent of the off-fault coseismic damage zone, i.e., the volume within the crust directly surrounding the rupture that experienced mechanical damage \cite{Bhat07, Templeton08, Thomas17_2, Thomas18, Okubo19, Okubo20}. Coseismic off-fault damage refers to fractures created or reactivated in the off-fault volume due to the dynamic rupture. As variations in rupture speed affect the seismic radiation and the off-fault coseismic damage, understanding the conditions for a rupture to transition from the sub-Rayleigh to the supershear regime would greatly help constrain fault properties (geometry, friction, lithology, etc.) and traction conditions that promote supershear ruptures. In addition, knowing how, why, and where earthquakes attain supershear speeds would help  in the reliable estimation of earthquake hazard assessment \cite{Madariaga83, Das07}. 

Whether supershear ruptures occur in nature was a matter of debate for a long time. While theoretical and numerical models demonstrated in the early 1970s that earthquakes could propagate at supershear speeds \cite{Burridge73, Hamano74, Andrews76, Das76, Das77, Burridge79, Freund79}, the absence of field observation and their extreme rarity in laboratory experiments \cite{Wu72} first suggested that supershear earthquakes could not exist in nature. It was not until the $M_{w}$~6.5 Imperial Valley earthquake (California, 1979) that a supershear rupture was inferred for the first time \cite{Olson82, Archuleta84,Spudich84}. Pioneering laboratory experiments \cite{Rosakis99, Xia04, Passelegue13} together with observations from the 1999 $M_{w}$~7.4 Izmit and the 1999 $M_{w}$~7.2 D\"uzce earthquakes in Turkey \cite{Bouchon00, Bouchon01}, then conclusively confirmed that supershear ruptures are more common than previously expected. Supershear ruptures have now been inferred for several, albeit rare, events: the 2001 $M_{w}$~7.8 Kunlun (China) earthquake \cite{Bouchon03_2, Robinson06, Vallee08}, the 2002 $M_{w}$~7.8 Denali (Alaska) earthquake \cite{Ellsworth04,Dunham04}, the 2010 $M_{w}$~6.9 Qinghai (China) earthquake \cite{Wang12}, the 2012 $M_{w}$~8.6 off-Sumatra (Indonesia) earthquake \cite{Wang12_1}, the 2013 $M_{w}$~7.5 Craig (Alaska) earthquake \cite{Yue13}, the 2013 $M_{w}$~6.7 Okhotsk (Kamtchatka) earthquake \cite{Zhan14, Zhan15}, and most recently the 2018 $M_{w}$~7.5 Palu (Indonesia) earthquake \cite{Bao19, Socquet19}.

Several efforts have been made to identify and characterize these supershear earthquakes using various methods \cite{Bouchon01, Bouchon08, Vallee08, Walker09, Bao19}. The methods are, in most cases, designed to reveal along which segment the rupture propagated at supershear velocities. The location of the transition from sub to supershear speeds is usually inferred as follows. The fault is subdivided into segments for which an average rupture speed is determined, using kinematic inversion methods. If the average rupture speed of a segment is deduced to be supershear, supershear transition is presumed to have occurred in between subshear and supershear segments. However, due to the lack of dense near-fault records, rupture speed estimates from seismological inversions and back-projection techniques rely on teleseismic records, leading to rough estimates of the transition's location (i.e., tens of kilometers precision). Even in a situation where near-field data exists, such as during the 2002 Denali earthquake \cite{Ellsworth04, Dunham04}, records are still too sparse to precisely locate the transition.

Numerous theoretical and numerical models have been developed to characterize the mechanics of supershear transition. On planar faults with homogeneous stress and strength, the occurrence of supershear ruptures depends on the $S$ ratio: $S  = (\tau_{p}-\tau_{0})/(\tau_{0}-\tau_{r})$, where $\tau_{0}$ is the background shear stress and $\tau_{p}$ and $\tau_{r}$ are the peak and residual frictional strengths \cite{Andrews76, Das77, Das76}. Supershear ruptures occur if $S < 1.77$ in 2D \cite{Das77, Andrews76}, or $S < 1.19$ in 3D \cite{Dunham07}, and the fault length is long enough for transition to occur \cite{Andrews76}. The time and location of the transition are determined by how far $S$ is from the Burridge-Andrews threshold of 1.77. However, natural faults do not have homogeneous stress and strength conditions before an earthquake. Numerical studies of supershear transition have also focused on spatial heterogeneity in stress and/or strength distribution, on a planar fault \cite{Dunham03, Dunham07, Liu08} and on a non-planar fault \cite{Shi13, Bruhat16, Hu16}, while Kaneko \& Lapusta \cite{Kaneko10} and Hu \textit{et al.} \cite{Hu19} looked at the effect of the free surface on supershear transition. 

Despite these efforts, the conditions for supershear transition in nature are still poorly understood because the imprecise location of the transition does not allow us to properly characterize the conditions on, or off, the fault that led to it. In addition, it also appears difficult to precisely capture this transition in both laboratory earthquakes (e.g., Mello \cite{Mello12}) and numerical models. This problem is mainly because the transition happens over very small length scales (of millimeters in the lab and over a few elements in numerical methods). A new approach that focuses specifically on the characterization of the region of transition is thus required to better understand why earthquakes accelerate to supershear speeds. 

Although the modeling efforts helped better understand the physics of supershear ruptures, it is not straightforward to apply that knowledge in the field, especially since we do not directly measure the stress and strength along the fault. Therefore, we propose a new approach based on the physical characterization of parameters that may promote supershear ruptures. This approach focuses on the detailed analysis of coseismic off-fault damage as a proxy to characterize the transition to supershear speeds. In this work, we cross-validate this idea by writing a theoretical analysis and comparing it with several numerical approaches. Then, we look at the observations available for natural supershear ruptures to corroborate this approach. 

\section{Coseismic Off-Fault Damage Around The Rupture Tip While Approaching Supershear Transition: Theoretical Analysis}
\label{sec:theory}

Coseismic off-fault damage depends on the evolution of stress during the propagation of a rupture in the crust. We use Linear Elastic Fracture Mechanics (LEFM) to provide closed-form solutions to quantify the state of stress around the rupture tip. 

Consider a semi-infinite plain-strain crack that moves at speed $v \le c_R$, where $c_R$ is the Rayleigh wave speed. Let the origin of the polar coordinate system $(r,\theta)$ coincide with the crack tip. The near-tip stress, which also depends on the rupture speed $v$, is given as \cite{Freund90},
\begin{eqnarray}
    \sigma_{\alpha\beta}(r,\theta,v) &=&  \frac{K^{dyn}_{II}(v)}{\sqrt{2\pi r}}f^{II}_{\alpha\beta}(\theta,v) + \sigma^0_{\alpha\beta} \quad \quad \alpha,\beta = 1,2\\
    \sigma_{33} &=& \nu(\sigma_{11} +\sigma_{22})
\end{eqnarray}
where $\sigma^0_{\alpha\beta}$ is the initial stress state and $K^{dyn}_{II}$ is the dynamic stress intensity factor which can be approximated as
\begin{equation}
   K^{dyn}_{II} \approx \frac{1 - v/c_R}{\sqrt{1 - v/c_p}} K_{II}^{sta}(\hat{L})
   \label{eqn:KII} 
\end{equation}
where $c_p$, the P-wave speed, is the limiting speed for a mode II crack and $\hat{L} = \int_{0}^{t} v(t)dt$ is the current crack length and $c_R$ is the Rayleigh wave speed \cite{Freund90}. Here $K_{II}^{sta}(L) = \Delta \tau\sqrt{\pi L}$ is the static stress intensity factor of the crack and $\Delta\tau$ is the stress drop defined as $\sigma^0_{yx} - \tau_r$. Here $\sigma^0_{yx}$ is the initial shear stress on the fault and $\tau_r$ is the residual frictional strength. This solution by Freund \cite{Freund72b} allows us to transform a static solution into a dynamic one. The rupture velocity dependence of the stress field makes it undergo a Lorentz-like contraction affecting both the stress field and its angular distribution as the rupture speed approaches the Rayleigh wave speed, $c_R$ \cite{Freund90}. This contraction has already been observed and verified experimentally for ruptures approaching the Rayleigh wave speed \cite{Svetlizky14}. For a crack-like rupture, there is a competing effect of the increase in stress intensity factor with increasing crack length. If it is a stable slip pulse of fixed length, then the effect of the crack length is invariant with rupture speed.

\begin{figure}[ht!]
\begin{center}
\includegraphics[width=\textwidth]{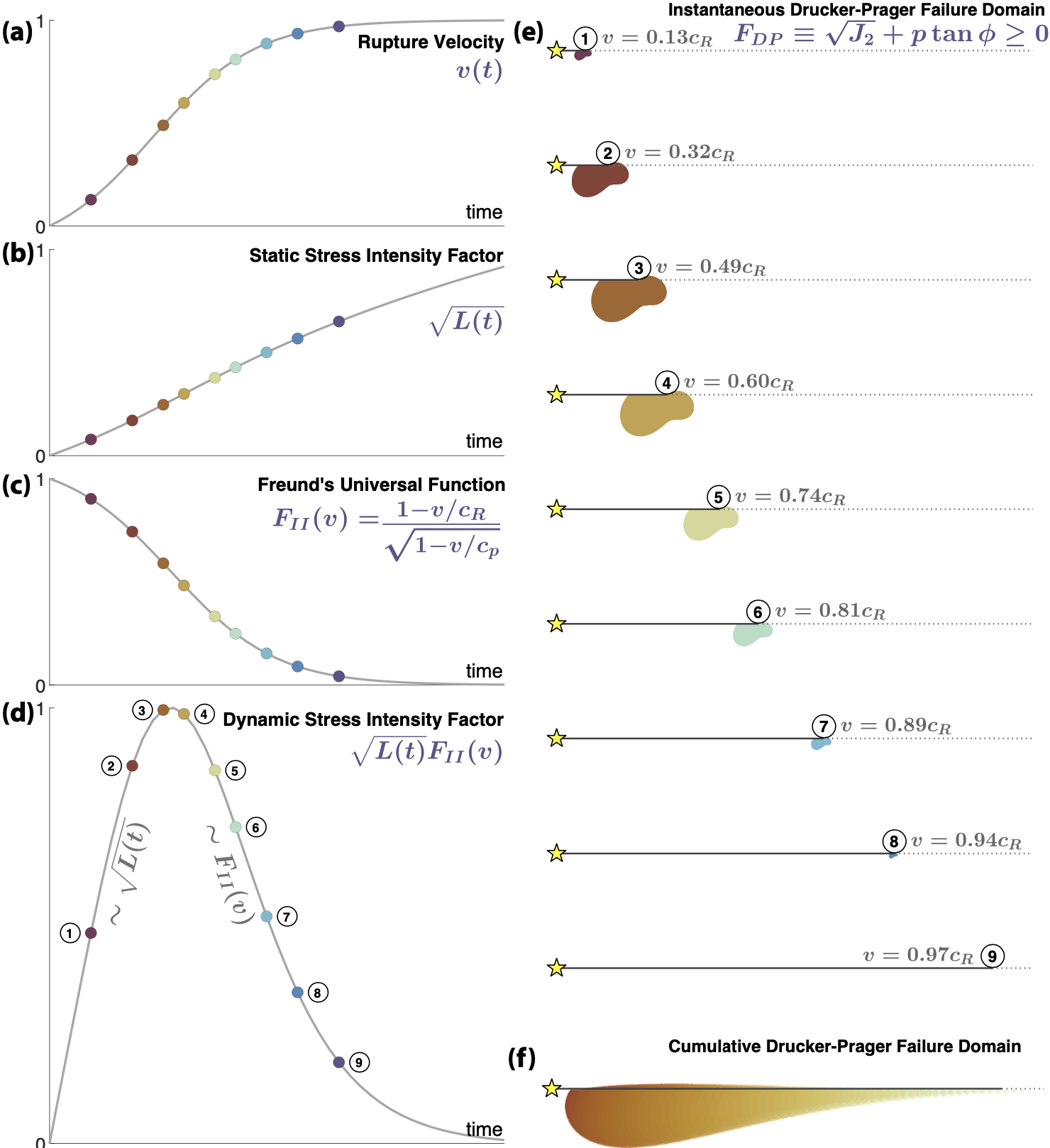}
\end{center}
\caption{\textbf{LEFM solution illustrating the temporal evolution of damage around a rupture tip}. All plots are normalized by their maximum values. \textbf{(a)} Prescribed rupture velocity history normalized by the Rayleigh wave speed. \textbf{(b)} Evolution of the static stress intensity factor which is proportional to the length of the crack, $\sqrt{L(t)}$. \textbf{(c)} Freund's universal function. Here $c_R$ and $c_p$ are the Rayleigh and P-wave speeds respectively. \textbf{(d)} Evolution of the dynamic stress intensity factor. \textbf{(e)} Instantaneous snapshots of the contours of the domain where the Drucker-Prager failure criterion is violated, $F_{DP} \ge 0$, at key points labeled 1 through 9 in \textbf{(d)}. \textbf{(f)} Superposition of the contours where $F_{DP} \ge 0$ throughout the history of the rupture.}
  \label{fig1a}
\end{figure}

When an earthquake rupture transitions to the supershear regime, the rupture first accelerates to the Rayleigh wave speed. Within the framework of LEFM, as the rupture approaches the Rayleigh wave speed, $K^{dyn}_{II}$ monotonically decreases to zero, strongly reducing the stress concentration at the rupture tip. Thus the off-fault domain affected by this stress concentration also shrinks. We illustrate this effect by calculating the extent of the region where the stress state exceeds the Drucker-Prager failure criterion, i. e. $F_{DP} \ge 0$, where
\begin{equation}
F_{DP} = \sqrt{J_2}+p\tan{\phi}.
\label{eqn:FDP}
\end{equation}
Here $p\equiv I_1/3=\sigma_{kk}/3$ is the hydrostatic stress derived from the first invariant of the stress tensor, $I_1$, $J_2=s_{ij}s_{ji}/2$ corresponds to the second invariant of the deviatoric stress tensor ($s_{ij} = \sigma_{ij} - p\delta_{ij}$) and $\tan{\phi} = f$ is the static coefficient of friction. This yield criterion, effectively a smooth approximation to the Mohr-Coulomb law using invariants of the stress tensor \cite{Drucker52}, takes into account all possible planes of slip, as opposed to Mohr-Coulomb where the potential slip planes need to be defined \textit{a priori} or are optimally oriented.


Using the LEFM solutions for a mode II crack, we can compute $F_{DP}$. The closed-form expressions are quite long, but the key variables that affect the solution can be written as
\begin{eqnarray}
    F_{DP} = \mathcal{F}\left[\dfrac{1}{\sqrt{r}},
    \sqrt{\hat{L}}\left(\dfrac{1 - v/c_R}{\sqrt{1 - v/c_p}}\right),\Delta\tau, \theta,\nu,\tan{\phi},\sigma^0_{\alpha\beta}\right]
\end{eqnarray}
Thus, given a rupture velocity history, $v(t)$, one can then calculate the region around the fault where the Drucker-Prager yield criterion is violated. 

We illustrate the solution in Figure \ref{fig1a}. Consider a fault in an elastic medium subject to an initial stress state (see Table 1 for all the parameters used). Assume that the rupture is accelerating to the Rayleigh wave speed (Figure \ref{fig1a}a). The instantaneous static stress intensity factor increases with the increasing length of the crack (Figure \ref{fig1a}b). However, Freund's universal function, $F_{II}(v)$, monotonically decreases to 0 (Figure \ref{fig1a}c). Consequently, the dynamic stress intensity factor first increases with increasing crack length but soon starts decreasing as it is dominated by $F_{II}(v)$ (Figure \ref{fig1a}d). Thus, we should expect to see an initial increase in the spatial extent of the damage followed by a reduction as the rupture approaches the limiting speed. We illustrate this effect at key moments in the rupture's history (marked by points numbered 1 through 9 in Figure \ref{fig1a}d). First, we compute, in Figure \ref{fig1a}e, the `instantaneous' damage, i. e., not accounting for the damage accumulated from the past history of the rupture around the rupture tip. We observe that the spatial extent of the domain where Drucker-Prager failure criteria is violated first increases (time points 1 through 3) and then decreases (time points 4 through 9) as the rupture accelerates. We can now superimpose all the snapshots of instantaneous damage to visualize the `cumulative' damage accumulated by the rupture throughout its history (Figure \ref{fig1a}f). Note that this last exercise is purely for illustrative purposes as the LEFM solution does not really account for the history of the rupture. These limitations will be overcome in the following sections. Nevertheless, to the first order, this approach illustrates that, as the rupture approaches the Rayleigh wave speed, the width of the coseismic off-fault damage zone should decrease significantly.

It is also worth noting that in the calculations above, the off-fault medium remains elastic. Thus, the rupture is insensitive to changes in the constitutive response of the medium due to coseismic off-fault damage. To resolve these issues, we need to take advantage of numerical simulations that allow for proper constitutive descriptions of the off-fault medium, including feedback from changes in elastic properties on the rupture behavior and that also allow for transition from subshear to supershear speeds.

\section{Coseismic Off-Fault Damage Around The Rupture Tip During Supershear Transition: Numerical Analysis}
\label{sec:NumericalModel}

While many efforts have been made to reveal the dynamics of supershear ruptures \cite{Dunham03, Dunham07, Bhat07, Liu08, Kaneko10, Shi13, Gabriel13}, the importance, during such event, of the complex feedback between the dynamic rupture and the mechanically evolving medium was not fully considered. 
However, new modeling approaches have recently emerged, allowing for dynamic activation of coseismic off-fault damage around faults and its feedback on rupture dynamics. In this section, we discuss two modeling strategies that allow for off-fault damage associated with brittle fracture. In the following, we show that the observed damage pattern does not depend on any specific constitutive law used to account for damage. Instead, crustal damage is a universal feature that can aid in identifying the region of supershear transition in the field.

\subsection{Modeling strategies}
\label{subsec:Model_strategy}

The two numerical models presented in this study account for fracture damage-dominated brittle rheology. Earthquake ruptures induce large dynamic strain rates ( $\ge 1s^{-1}$) around the rupture tip that cannot be fully accounted for by classical Mohr-Coulomb or Drucker-Prager plasticity, typically used low strain rates ( $\le 10^{-5}s^{-1}$)  and the two models presented below allow to account for such large dynamic strain.
Okubo \textit{et al.} \cite{Okubo19, Okubo20} explicitly model the sub-kilometric off-fault fractures using a continuum-discontinuum numerical framework. The micromechanical model \cite{Thomas17_2, Thomas18} accounts for coseismically activated ``microfractures'' ($\sim$10's of meters) by computing the dynamic changes in the constitutive response due to these fractures. The difference between the two models is essentially in the modeled off-fault fracture networks scale, leading to two distinct modeling strategies. In essence, the two brittle failure models are complementary, and they should, ideally, be combined. Nevertheless, they are good proxies for modeling off-fault damage at various scales, and we further show that they produce similar results.

To reproduce coseismic damage at relatively large scales (i.e., 50m to several km), Okubo \textit{et al.} \cite{Okubo19, Okubo20} use the software suite HOSSedu, developed by the Los Alamos National Laboratory (LANL) \cite{Rougier16, Knight20}. The numerical algorithms behind this tool are based on the combined Finite-Discrete Element Method (FDEM) proposed by Munjiza \cite{Munjiza04}, to produce dynamically activated off-fault fracture networks. The key feature here is that FDEM allows each interface between the finite elements, describing the off-fault medium, to have its own tensile and shear failure criterion. Thus, these interfaces can rupture under appropriate traction conditions. When the earthquake rupture propagates, the dynamic stress field around its tip will increase to violate the tensile or shear failure criteria leading to off-fault fracture damage. The lower limit of fracture resolution is around 50m, which is defined by the minimum size of the discretized mesh. For a further detailed description of the method, see Okubo \textit{et al.} \cite{Okubo19, Okubo20}.

The second modeling strategy relies on laboratory experiments \cite{Walsh65a,Walsh65b,Faulkner06} and field observations \cite{Hiramatsu05,Froment14} that show significant changes in elastic properties related to fracturing. Observations in the field show up to 40\% coseismic reduction in P- and S-wave velocities on spatial scales of hundreds of meters normal to the fault and few kilometers in depth. Following the damage constitutive laws proposed by Ashby \& Sammis \cite{Ashby90}, Deshpande and Evans \cite{Deshpande08}, Bhat \textit{et al.} \cite{Bhat12}, Thomas \textit{et al.} \cite{Thomas17_2} implemented an energy-based micromechanical brittle rheology, as developed by Rice \cite{Rice75}, to account for such dynamic change of bulk rheological properties during earthquakes. 
In short, at each equilibrium state, the Gibbs free energy density $\Psi$ of the damaged solid is defined as the sum of (1) the free energy $\Psi^{e}$ of a solid, without flaws, deforming purely elastically and (2) the free energy $\Psi^{i}$ corresponding to the contribution of the current set of microcracks. Using thermodynamic arguments, $\Psi = \Psi^{e}+\Psi^{i}$ is used to derive the new stress-strain constitutive law and the changes of elastic properties in the medium at the equilibrium stage. In the model, following laboratory experiments,  the evolution of $\Psi^{i}$ is determined by taking into account the effect of loading rate and crack-tip velocities on crack growth (see Zhang \& Zhao \cite{Zhang13} for an overview). A complete description of the model can be found in Bhat \textit{et al.} \cite{Bhat12} and Thomas \textit{et al.} \cite{Thomas17_2}. 

\subsection{Model set up}
\label{subsec:setup}

For comparison purposes, we set up the two brittle rheology models to be as close to each other as possible. In both cases, we consider a 2-D plane strain medium, with a 1-D right lateral fault embedded in it and loaded by uniform background stresses. The maximum compressive stress $\sigma_1$,  and the minimum compressive stress $\sigma_3$  are in the $x-y$ plane, whereas the  intermediate principal stress $\sigma_2$ coincides with $\sigma_{zz}$. The fault plane makes an angle of $60^o$ with $\sigma_1$ with normal uniform traction ($\sigma_{yy}^0$) and shear traction ($\sigma_{xy}^0$) everywhere except in the nucleation zone of the micromechanical model where the shear traction is slightly above the nominal static strength. In the FDEM model, the dynamic rupture is initiated by locally decreasing the static friction, in the nucleation zone, instead of the local change of $\sigma_{xy}^0$. These different nucleation strategies do not affect the results since this study is focused on damage occurring when the rupture is dynamic (sub and supershear). For each model, the initial shear stress and the $S$-ratio were chosen so that the rupture transitions to supershear speed reasonably early on.

Rupture propagation along the main fault plane is governed by a slip-weakening friction law \cite{Palmer73}. Static friction is set at 0.6, which corresponds to a value measured in laboratory experiments for a large range of rocks \cite{Byerlee78}, and dynamic friction at 0.1 as observed in high slip-rate experiments \cite{Wibberley08}. 

The fault length is 64km for the micromechanical model and 115km for the FDEM model. The domain width is determined based on the numerical method and on the scale of the fracture networks accounted for in the calculations. For the micromechanical model, the width around the fault is 6km with absorbing boundary conditions to avoid reflections interfering with the propagating dynamic rupture. For the FDEM model, we simply set the domain large enough (86km) so that the reflections do not arrive on the fault over the computation duration.
\renewcommand{\arraystretch}{0.75}
\begin{table}[ht!]
\caption{Table of parameters for the LEFM, Micromechanical and FDEM models}
\begin{center}
\begin{tabular}{c l l}
\hline
Parameters & Description & Values\\
\hline
\multicolumn{3}{l}{\textbf{State of stress}}\\
$-\sigma_{xx}^0$ & Fault-parallel stress (MPa) & 37.7\\
$-\sigma_{yy}^0$ & Normal stress on the fault (MPa)  & 60.7 \\
$\sigma_{yx}^0$ & Shear stress on the fault (MPa)& 19.9\\
 & Nucleation Zone$^{*}$ (MPa) &36.4\\
$\psi$ & Angle between $\sigma_1$ and fault (deg.) & 60 \\
\multicolumn{3}{l}{\textbf{Bulk properties}} \\
 $\rho$ & density (kg/m$^3$)	&2700  \\
 $c_s$ & S-wave speed	(km/s)& 3.12 \\
$c_p$ & P-wave speed	(km/s)&	5.62\\
$\nu$ & Poisson's ratio &0.276 \\
\multicolumn{3}{l}{\textbf{Fault parameters}}\\
$f_s$ &Static friction coefficient & 0.6 \\
 & Nucleation Zone$^{\dagger}$ & 0.3\\
$f_d$ &Dynamic friction coefficient & 0.1 \\
$D_c$ & Characteristic slip distance (m) &  1.0\\
$R_0$ & Process zone size	(m)& 1057\\
$S$ & $S$-ratio & 1.2\\
\hline
\end{tabular}
\label{tab:model_param}
\end{center}
\end{table}
We non-dimensionalize all length scales by the static size of the process zone ($R_0$) so that the models could be compared to natural or laboratory earthquakes. $R_0$ is defined as
\begin{equation}
\label{eqn:pro_zone}
    R_0 = \frac{9\pi }{16(1-\nu)}\frac{\mu G}{(\tau_p-\tau_r)^2}.
\end{equation}
Here $\nu$ is Poisson's ratio, $\mu$ is shear modulus, $G$ is the fracture energy associated with the slip-weakening law, $\tau_p$ and $\tau_r$ are the static and dynamic frictional strengths, respectively.

Reference values for the common parameters between the two models are summarized in Table \ref{tab:model_param}, whereas parameters specific to different modeling strategies are provided in the Supplementary Information (Tables S1 \& S2).  Schematics and parameters for the simulations can also be found in Figure S1.
 
\subsection{Influence of rupture dynamics on off-fault damage}
\label{subsec:modelresult}

\begin{figure}[ht!]
\begin{center}
\includegraphics[width=0.8\textwidth]{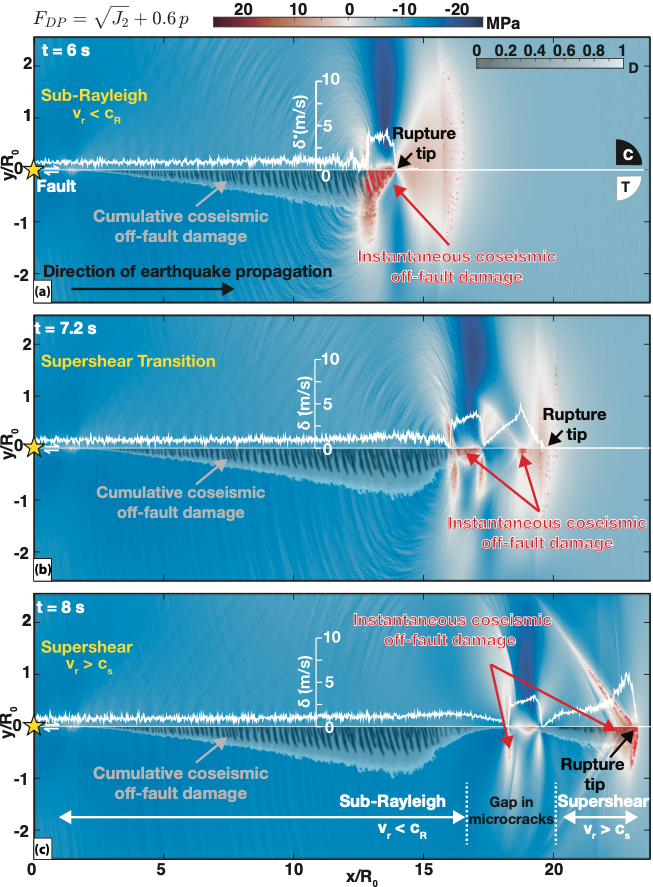}
\end{center}
\caption{\textbf{Damage and Mohr-Coulomb Failure with the micromechanical model}. Temporal evolution  of a dynamic rupture occurring on a right-lateral fault embedded in granite (solid grey line at $y=0$), using the micromechanical model. The rupture is nucleated around $(0,0)$ denoted by the yellow star. Sub-figures display snapshots of the Drucker-Prager yield criterion (equation \ref{eqn:FDP}) when the rupture propagates at subshear \textbf{(a)} and supershear velocities \textbf{(c)}, and during the transition \textbf{(b)}.  Fault slip rate (white curves) and the cumulative damage density, $D$, (grey scale) are superimposed. Instant damage happening in relation to the displayed stress field is underlined in red. Yellow star denotes the nucleation patch and C and T the compressional and the tensional quadrants, respectively.}
  \label{fig:damage_story}
\end{figure}

To understand the complex interaction between the main fault and the surrounding medium undergoing coseismic damage, it is necessary to simultaneously look at the rupture dynamics, the associated stress field around the fault, and the triggered damage in bulk. Here we use the micromechanical model to underline the key features that arise from numerical studies. We note that the same conclusions can be drawn from the FDEM model.

\begin{figure*}[ht!]
\begin{center}
\includegraphics[width=0.95\textwidth]{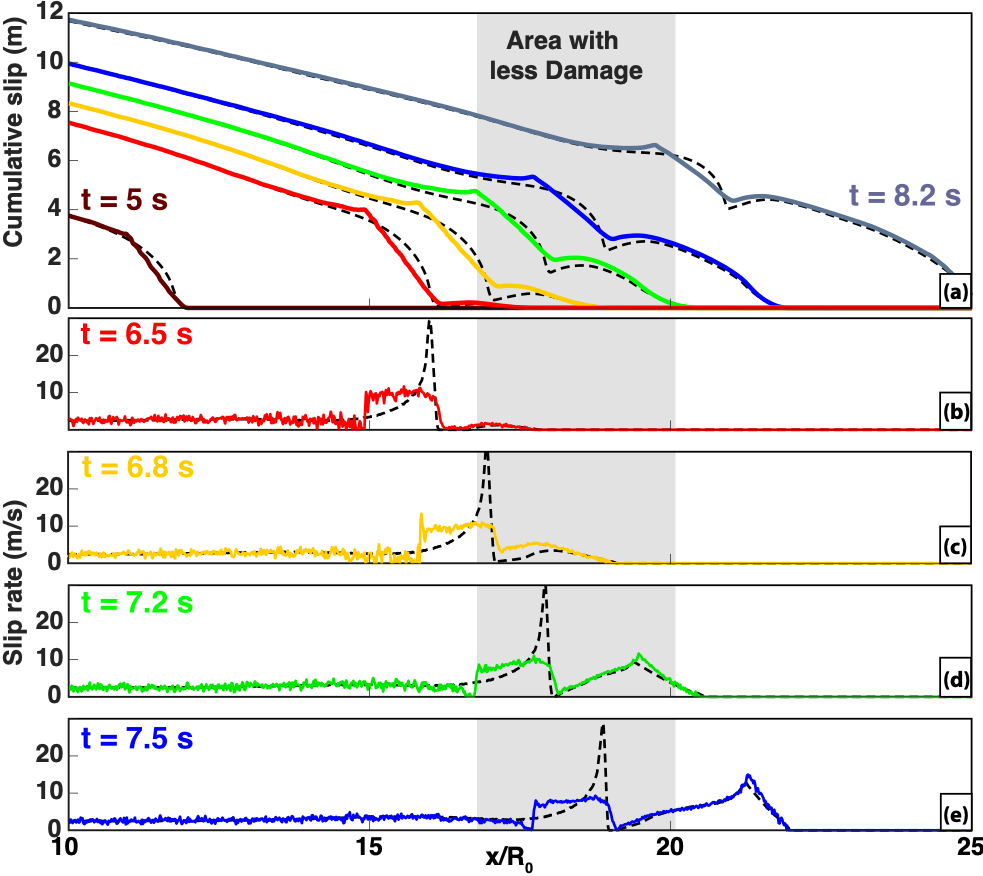}
\end{center}
 \caption{\textbf{Slip and slip rate evolution with the micromechanical model}. Cumulative slip \textbf{(a)}, and slip rate, $\dot{\delta}$ , \textbf{(b-e)} on the fault are displayed at various time steps. Colored curves correspond to the dynamic simulation with the damage evolution law, dotted grey curves represent a simulation with the same parameters but within a pure elastic medium. Grey box gives the location of the area where less damage is observed. }
\label{fig:slipsliprate}
\end{figure*}

In the micromechanical model, under a compressive regime, damage occurs by the growth of pre-existing ``flaws''. They represent the faults, joints, cracks,  mineral twins, defects in the crystal structure, grain boundaries, etc., we observe in nature. 
Frictional sliding occurs on these pre-existing fractures when the shear stress overcomes the frictional resistance acting on the fracture interface.
As the faces slide in opposing directions, it creates a tensile wedging force that opens wing cracks at the tips of the shear fracture. The wing cracks grow when their stress intensity factors overcome the fracture toughness. In Figure \ref{fig:damage_story} we display the Drucker-Prager criterion to emphasize the regions where shear sliding is likely to occur, i.e., $F_{DP}>0$ (equation \ref{eqn:FDP}), and therefore where we expect wing cracks to grow. Fault slip rate (white curves) and cumulative damage (greyscale) are superimposed on each snapshot of $F_{DP}$. New coseismic damage induced by the displayed stress field is highlighted in red.
%
As the rupture propagates below the Rayleigh wave speed (Figure \ref{fig:damage_story}a), the damage is essentially generated behind the rupture front, in the tensional quadrant. This feature has been seen by Okubo \textit{et al.} \cite{Okubo19} as well, using the FDEM model. Damage also occurs ahead of the rupture front, where the S-wave field concentrates, although the resulting damage density is much smaller than behind the rupture front. The location of newly formed cracks corresponds to the region where $F_{DP}$ is positive.
%
The transition to supershear velocity (Figure \ref{fig:damage_story}b) directly impacts the generation of coseismic damage. Around $t=6.5$s a new pulse is generated ahead of the rupture (Figure \ref{fig:slipsliprate}b). The transition to supershear velocity coincides with a decrease in the width of the damage zone (Figures \ref{fig:damage_story}b and \ref{fig:slipsliprate}b-d). As we argued in section \ref{sec:theory}, this is likely related to the decrease in the stress intensity factor as the rupture speed increases.  When approaching the Rayleigh wave speed, $K_{II}^{dyn}$ becomes small, even if the total length of the fault that has ruptured ($\hat{L}$ in equation \ref{eqn:KII}), has increased significantly. 
Moreover, a snapshot of $F_{DP}$ at $t=7.2$s (Figure \ref{fig:damage_story}b) shows that the first pulse is not strong enough to change the state of stress optimally, thus reducing even further $K_{II}^{dyn}$. These two factors likely explain why damage mostly occurs behind the front of the pulse propagating at sub-Rayleigh rupture speed.
%
After $t=7.2$s, the second pulse ahead of the rupture front becomes stronger and the first pulse weaker (Figure \ref{fig:slipsliprate}d-e). Consequently, damage now essentially occurs behind the rupture front propagating at supershear speed and in the area behind the Mach cone  (Figure \ref{fig:damage_story}c). Therefore, even if the sub-Rayleigh pulse propagates now inside the ``transition zone", it induces little to no further damage. Therefore, this local reduction in the damage zone width is expected to remain unchanged for the earthquake duration and could, in principle, be observed in the field.

Finally, it is worth noticing that while the surrounding medium records the transition to supershear rupture (via the damage zone width), the final displacement along the fault, that geodesy could provide, carries no such information. As one can see in Figure \ref{fig:slipsliprate}a, the slip continues to accumulate in the ``transition zone" after the passage of the rupture, while the damage zone width remains unchanged (Figure \ref{fig:damage_story}c).

\subsection{Comparison between the different modeling strategies}
\label{subsec:model_comparison}

We now perform a similar exercise using the FDEM model and choose parameters such that the micromechanical and FDEM models are as close to each other as possible. 
We adopt a slightly different yield criterion to reflect on the way damage is accounted for in this numerical method (see section \ref{subsec:Model_strategy} and Okubo \textit{et al.} \cite{Okubo19}). 
We evaluate the potential regions of failure, in shear, by calculating the invariant form of the Mohr-Coulomb yield function, $F_{MC}$ (see equation 4.142 of Chen \& Han \cite{Chen07}):  
\begin{equation}
F_{MC} = R_{MC}\sqrt{J_2} + p\tan{\phi} - c
\label{eqn:FMC}
\end{equation}
where $c$ is the cohesion and $R_{MC} $ is given by:  
\begin{equation}
R_{MC} =\sin{\left( \Theta + \dfrac{\pi}{3}\right)} \sec{\phi}+ \dfrac{1}{\sqrt{3}}\cos{\left( \Theta + \dfrac{\pi}{3}\right)}\tan{\phi}
\end{equation}
where
\begin{equation}
\cos{3\Theta} = \dfrac{3\sqrt{3}}{2}\dfrac{J_{3}}{J_{2}^{3/2}}
\quad\mathrm{;}\quad
J_{3} = \dfrac{1}{3}s_{ij}s_{jk}s_{ki}
\end{equation}
where $J_3$ is the third invariant of the deviatoric stress tensor.
Failure occurs when $F_{MC} \geq 0$. Note that when $R_{MC} = 1$ and $c=0$, $F_{MC}$ becomes the Drucker-Prager criterion as described in equation \ref{eqn:FDP}.

\begin{figure}[ht!]
\includegraphics[width=\textwidth]{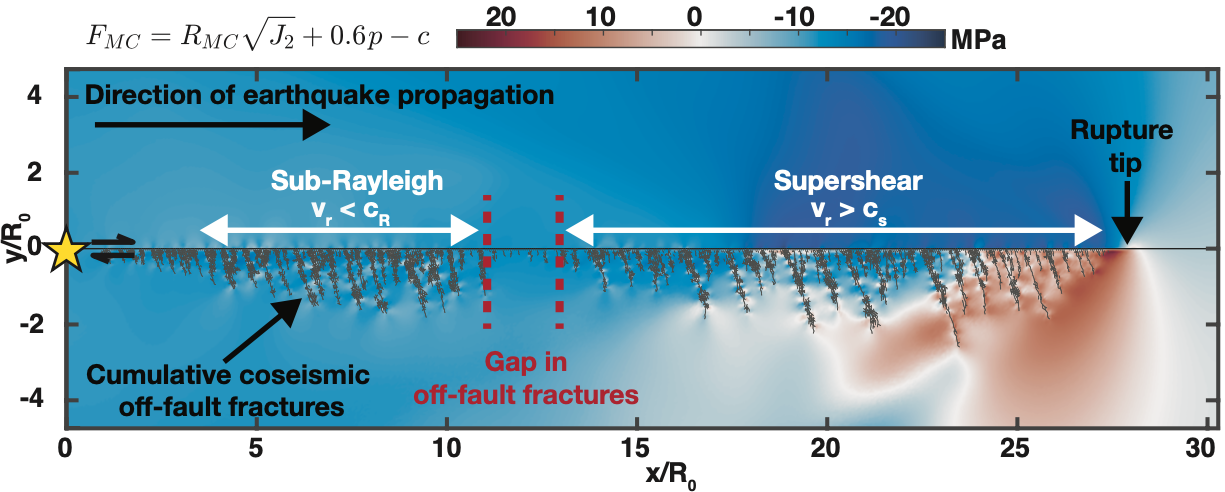}
\caption{\textbf{Damage and Mohr-Coulomb Failure with FDEM model}. Dynamic rupture occurring on a right-lateral fault embedded in granite, using the FDEM model. Figure displays the invariant form of the Mohr-Coulomb yield criterion (equation \ref{eqn:FMC}) when the rupture propagates at supershear velocities. Black lines gives the spatial distribution of off-fault fractures that occurs during the entire event. Signature of the transition from the sub-Rayleigh to the supershear regime is highlighted by the gap in off-fault damage.}
\label{fig:FDEMsimu}
\end{figure}

Figure \ref{fig:FDEMsimu} displays the $F_{MC} $ criteria when the rupture propagates at supershear velocities and the cumulative damage pattern resulting from the entire coseismic rupture. 
On top of the area undergoing $F_{MC}>0$ in relation to the rupture propagation along the main fault plane, we observe positive values at the tip of the discretized off-fault fractures, far away from the rupture front.  This feature is not so evident within the micromechanical model because the fractures are homogenized in the constitutive law (section \ref{subsec:Model_strategy}). We also observe a shrinkage in the spatial extent of damage when the rupture transitions to supershear, as seen earlier with the micromechanical model.

In addition to the above described numerical experiments using brittle rheology, we acknowledge that Templeton \& Rice \textit{et al.} \cite{Templeton08} first explored the extent and distribution of off-fault plasticity during supershear rupture (see Figure 13 of Templeton \& Rice \textit{et al.} \cite{Templeton08}). They conducted a 2D in-plane dynamic rupture modeling with Drucker-Prager elasto-plastic constitutive law. In their model, the degree of damage is inferred from the accumulation of plastic strain. Although plasticity, as a proxy for damage, essentially holds only at low-strain rates \cite{Davis05, Zhang13}, they nevertheless pointed out a ``remarkable contraction'' in the damage zone width associated with supershear transition. 

We can thus conclude, quite confidently, that this reduction in damage zone width (a gap in off-fault fractures) is a universal characteristic of supershear transition and is insensitive to the constitutive law used to model damage. 

\section{Natural observations of coseismic off-fault damage}
\label{sec:observations}

Theoretical and numerical models, with three different rheological descriptions of off-fault damage \cite{Templeton08, Thomas17_2, Okubo19}, all suggest that the region affected by the stress field around the rupture tip shrinks during the supershear transition. This shrinkage of the stress field results in a narrow off-fault damage zone. We now look for a direct (or indirect) signature of this reduced damage zone in the case of natural supershear earthquakes, verifying their location with corresponding kinematic models that invert for rupture speeds.

Vallage \textit{et al.} \cite{Vallage15} and Klinger \textit{et al.} \cite{Klinger18} showed that there is a one-to-one relationship between the features of the displacement field, around the fault, of an earthquake and off-fault fracture damage. Using the FDEM method described earlier, Klinger \textit{et al.} \cite{Klinger18} showed that the observed, spatially diffuse, displacement field around a fault is due to displacement accommodated by off-fault fractures (see Figure 3 in \cite{Klinger18}). This feature is in stark contrast with the sharp displacement field obtained, around a fault, when displacement is only accommodated by the main fault. Thus deviations from this sharp displacement field would allow us to characterize the width of the damage zone around a fault. 

We can also make a reasonable assumption that the newly created/reactivated off-fault fractures are likely to host early off-fault aftershocks (based on the observed stress field). We should then expect a reduction in the spatial extent of the early off-fault aftershocks at the location of the supershear transition. A more thorough analysis would involve continuing the simulations described earlier for up to a week, or so, after the earthquake. However, computational limitations make this beyond the scope of the present study.

\subsection{Optical image correlation to observe off-fault coseismic damage}

Recent developments in satellite optical image analysis and sub-pixel correlation methods allow for detecting displacement variations due to an earthquake down to sub-metric resolutions (e.g., Vallage \textit{et al.} \cite{Vallage15}, Barnhart \textit{et al.} \cite{Barnhart20}, Delorme \textit{et al.} \cite{Delorme20} and Ajorlou \textit{et al.} \cite{Ajorlou21}). These methods enable characterizing the surface rupture geometry, the amount of surface displacement, and the width of the zone affected by this displacement after an earthquake, also referred to as the ``fault zone width'' \cite{Milliner15}. The fault zone width  reflects the lateral extent of rock damage, and secondary deformation features around the fault \cite{Rockwell02,Mitchell09}. In the field, the fault zone width corresponds to the last deformation features observed when moving away from the fault core \cite{Rockwell02}. The same definition is employed in geodetic studies providing a more detailed characterization of this region due to the density and resolution of the observations  \cite{Klinger18}. 

The 2001 $M_{w}$~7.8 Kunlun (China) earthquake is a strike-slip event with $\sim$~400km long surface rupture and a mean rupture speed between 3.3-3.9~km/s, which is larger than the shear wave speed and hence reported as a supershear rupture \cite{Bouchon03_2,Robinson06}. This earthquake was well recorded by satellite images, allowing us to investigate the spatial distribution of coseismic off-fault damage by focusing on the fault zone width. We use SPOT-1 to SPOT-4 images, covering the 2001 Kunlun fault area from 1988 to 2004 with a ground resolution of 10m, allowing for change detection of less than 1m \cite{Vallage15, Rupnik17}. We focus on the central part of the rupture dominated by fault parallel motion \cite{Klinger05,Klinger06}, obtaining through optical image correlation the horizontal displacement field for the earthquake east-west and north-south components (Figure S2). The maximum strike-slip displacement for this area of the rupture is around 8 m \cite{Klinger05,Klinger06}.

\begin{figure}[ht!]
\begin{center}
\includegraphics[width=1\textwidth]{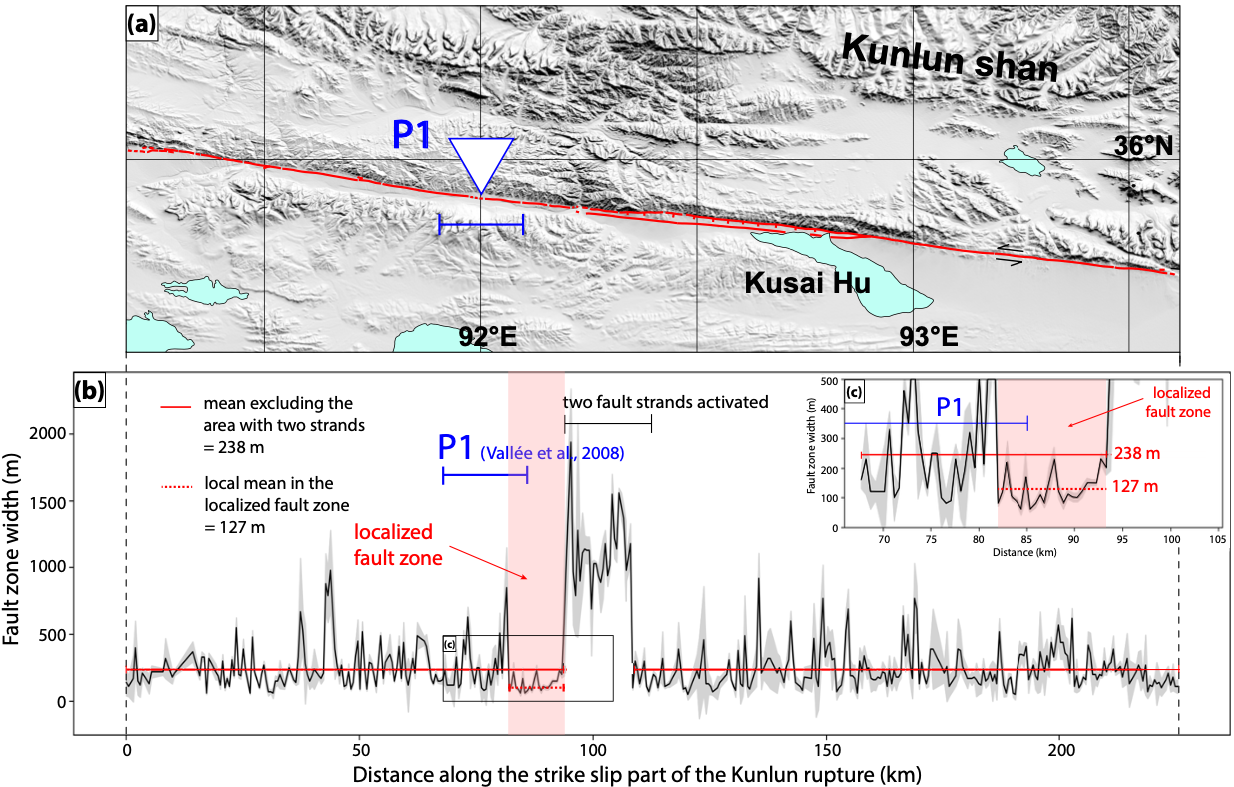}
\end{center}
\caption{\textbf{Optical Image Correlation Analysis}. \textbf{(a)} Map of the strike slip section of the 2001 $M_{w}$~7.8 Kunlun earthquake (China), where  P1 denotes the transition zone reported for the event from seismological far-field data \cite{Vallee08}. \textbf{(b)} Along-Strike fault zone width (black) and its associated uncertainty (grey), obtained from the analysis of 40km long profiles, sampling the fault zone every 500m, on the surface displacement maps. The latter is derived from correlating pre- and post-earthquake SPOT-1 to SPOT-4 images. The 11km long red area highlights a region with a mean fault width (red dashed line) of only 127m compared to 238m recorded for the rest of the rupture (red line). The latter excludes the area where two parallel fault strands are activated and for which the fault zone is exceptionally large (> 1000m). \textbf{(c)} Zoom of the Figure \ref{fig3}b.}
\label{fig3}
\end{figure}

Prior to correlation,  SPOT images are corrected from viewing angle and topography using a unique 90m SRTM digital surface model. The pre-and post-earthquake corrected paired images are correlated using the MicMac package \cite{Rosu15, Rupnik17}, which measures the horizontal surface displacement between the acquisitions (see Supplementary Information, Figure S2 to see the data employed in this study). The MicMac method enables preserving the resolution of the input images, that is 10m, since it measures the displacement at every pixel location in the pre-earthquake image. The size of the pixel pattern considered in the pre-earthquake images, and that is searched for in the post-earthquake ones, is 5 pixels, so 50m. A regularization of 0.3 is used to force MicMac to consider realistic displacement values relative to neighboring pixels, even if this procedure assumes pixels with a lower correlation score (regularization 0 in Micmac means taking the best correlation score without regarding on the coherence of surrounding pixels). Output displacement maps show displacement in the east-west and north-south directions. We measure 460 stacked profiles perpendicular to the fault trace in the displacement map (Figure S2) every 500 m to analyze variations in the surface displacements and fault zone width. Individual slip profiles are stacked on rectangles of 40km-long and 1km-wide, in order to average out noise and artifacts (Stacking profiles from ENVI version 5.5.1; Exelis Visual Information Solutions, Boulder, Colorado). On each profile, displacement is estimated in the fault-parallel direction, observing the coseismic offset produced by the earthquake. (see Supplementary Information, Figure S3 for a profile example). This procedure allows us to evaluate the fault zone width in the strike-slip direction, over 396 profiles (Fig.~\ref{fig3}b). Field measurements of fault zone widths are at least a few tens to a few hundred meters in the case of this earthquake \cite{Klinger05, Klinger06}. Thus, we infer that the fault zone width can be estimated using displacement maps, employing satellite image correlation. 

The analysis of the profiles highlights three main domains in fault zone width. The first domain, which represents the majority of the rupture, exhibits a mean value of 238~$\pm$~80m, and it corresponds to the strike-slip section of the rupture where only one fault zone is observed at the surface (Figures \ref{fig3}, S4A , S5). The second domain presents a width larger than 1km lasting for about 15km. This section shows a complex rupture pattern, where the rupture splits into two fault zones evolving into slip-partitioning between pure strike-slip along the southern fault strand and pure normal faulting along the northern strand. The slip-partitioning lasts for about 70 km eastward \cite{King05,Klinger05} (Figures \ref{fig3}, S4B, S5). The third domain is located right before the onset of the slip-partitioning region, where the average fault zone width drops to 127~$\pm$~39m over ~11km long segment. This value is about half of the mean fault zone width of 238m calculated for the whole fault zone (ignoring the slip-partitioning area). This area co-locates with the region where the rupture is inferred, from teleseismic data, to have transitioned to the supershear regime \cite{Bouchon03_2,Robinson06,Vallee08}. The onset of slip-partitioning cannot cause a reduction in the damage zone width because the normal faulting motion was quite likely triggered by the dynamic stress field of the main supershear segment \cite{Bhat07}. This means that the overall dynamics in that region was still dominated by the rupture on the strike-slip segment. With these observations, and considering the theoretical analysis and numerical modeling conducted in the previous sections, we interpret this localized fault zone width reduction as a shrinkage of the off-fault damage zone associated with the rupture transitioning to supershear speeds (e.g., Figure \ref{fig:damage_story}). Hence, this along-strike feature seen by the optical correlation analysis represents a natural observation of the supershear transition zone.

\subsection{Distribution of early aftershocks as a proxy for off-fault coseismic damage}
\label{sec:aftershocks}

Assuming that the nucleation of early aftershocks is mainly governed by the stress state left in the wake of the earthquake, we test whether the extent of coseismic off-fault damage in nature manifests in the spatial distribution of early off-fault aftershocks. The logic is that considering the state of stress in the wake of the earthquake, weakened regions will preferentially host early aftershocks, and thus we should observe a region with less events where the rupture transitioned to supershear. When re-analyzing the regions where established supershear ruptures transitioned from sub-Rayleigh to supershear speeds, we might find local minima in the spatial extent of early aftershocks. If this region coincides with kinematic observations of supershear transition, we would have an independent and more precise location of supershear transition.

\begin{figure}[ht!]
\begin{center}
\includegraphics[width=1.\textwidth]{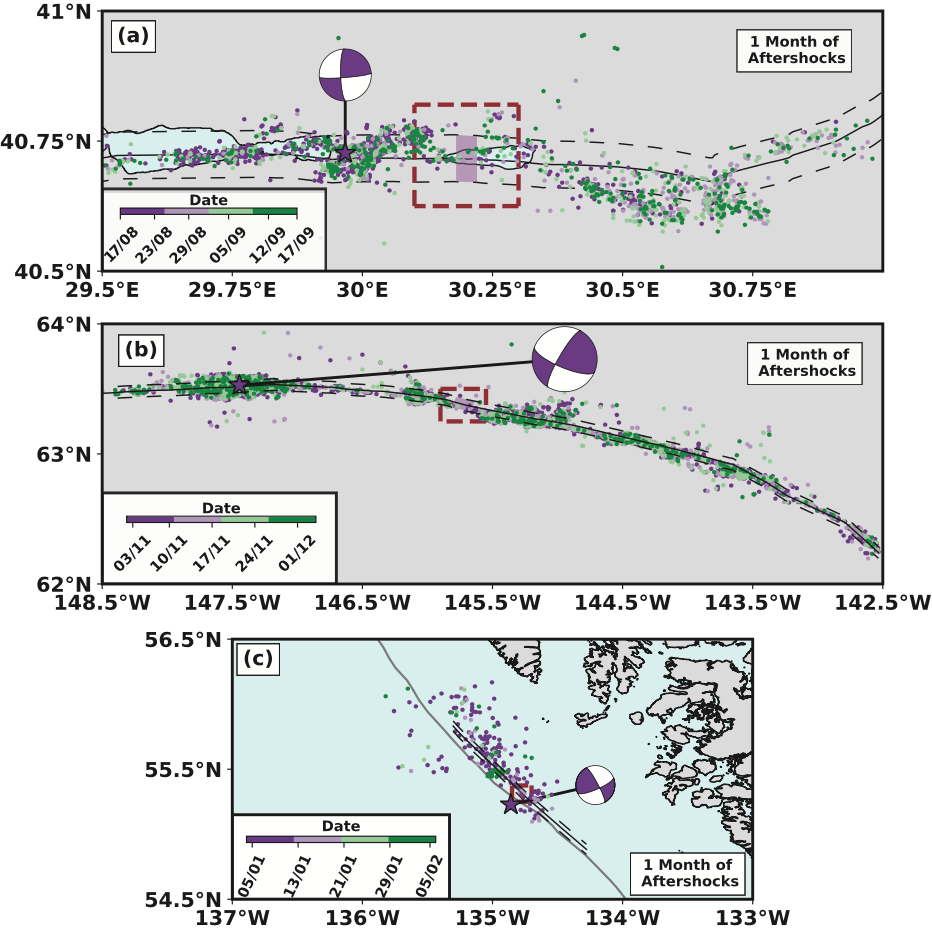}
\end{center}
  \caption{\textbf{Aftershock Catalogs}. 1-month aftershock distribution for Izmit (\textbf{a}), Denali (\textbf{b}), and Craig (\textbf{c}) earthquakes, color-coded by time. Pink stars indicate the mainshock epicenters with focal mechanism. The continuous black lines denote the surface rupture for each event, while the dashed ones indicate a distance of 5km from the fault for each event. Brown dashed boxes are the target regions to explore the supershear transition based on published kinematic models (Bouchon \textit{et al.} \cite{Bouchon00} for \textbf{a}, Ellsworth \textit{et al.} \cite{Ellsworth04} for \textbf{b}, and Yue \textit{et al.} \cite{Yue13} for \textbf{c}. The pink boxes indicate this work’s proposed transition zones. }
  \label{fig2a}
\end{figure}

We analyze three reported supershear ruptures for which high-resolution aftershock catalogs are available, at least up to one month after the main event. The 1999 $M_{w}$~7.4 Izmit earthquake (Turkey) had a mean rupture velocity of about 4.8km/s \cite{Bouchon00,Bouchon01}, and a well-recorded 3-month aftershock catalog \cite{Bouchon08} (Figure \ref{fig2a}a for 1-month aftershock locations). In the case of the 2002 $M_{w}$ 7.9 Denali earthquake (Alaska), rupture velocities of about 5.5km/s were reported \cite{Ellsworth04}, and 1-year long catalog of aftershocks is available \cite{Ratchkovski03} (Figure \ref{fig2a}b for 1-month aftershock locations). The 2013 $M_{w}$ 7.5 Craig earthquake (Alaska) was also inferred as a supershear event, with rupture velocities between 5.5 to 6.0km/s \cite{Yue13} and a 5-month aftershock catalog is available \cite{Holtkamp15} (Figure \ref{fig2a}c for 1-month aftershock locations). 

\begin{figure}[ht!]
\begin{center}
\includegraphics[width=0.7\textwidth]{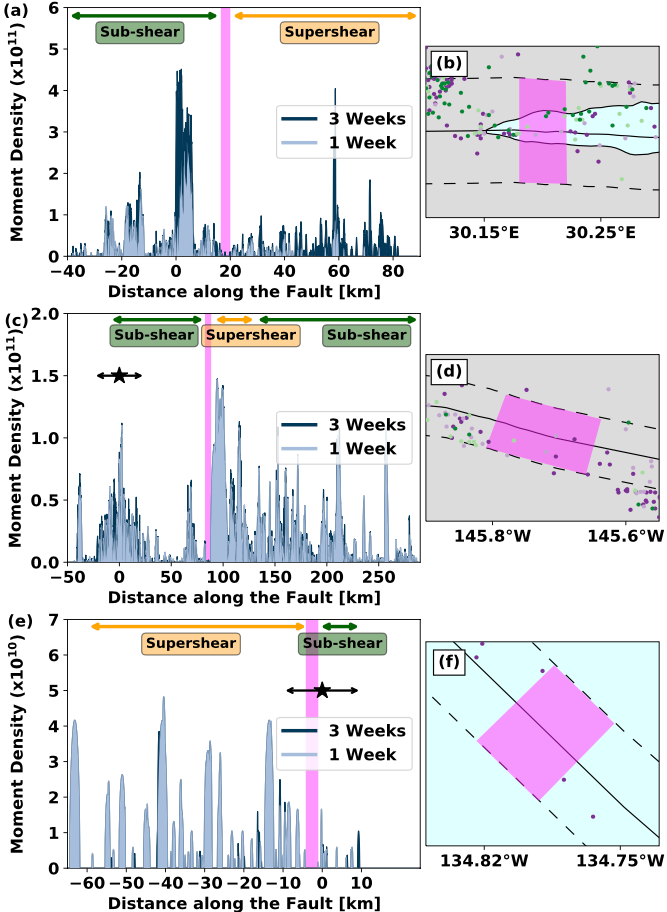}
\end{center}
  \caption{\textbf{Spatio-Temporal Seismic Moment Density Evolution}. Cumulative seismic moment density projected on the main fault at different temporal scales (1-3 weeks), for Izmit (\textbf{a}), Denali (\textbf{c}), and Craig (\textbf{e}) earthquakes. All the aftershocks within a distance of 5~km from the fault are considered in the calculation (area denoted by the black discontinuous lines in Figures \ref{fig2a}a, \b and c). Color-coded arrows (at the top in \textbf{a}, \textbf{b}, and \textbf{c}) indicate the different speed regimes reported for each event (green for sub-Rayleigh and orange for supershear) \cite{Bouchon08,Ellsworth04,Yue13}, while the starts denote the epicenter of each earthquake and the arrows indicate the direction of the rupture. The pink boxes highlight our proposed transition zone, also observed in a map view in (\textbf{b}) for Izmit, (\textbf{d}) for Denali, and (\textbf{f}) for Craig earthquakes.}
  \label{fig2b}
\end{figure}

We observe a paucity in aftershocks where the ruptures have transitioned to supershear, as highlighted by the pink boxes in Figures \ref{fig2b}b, d, and f. We quantify such lack of aftershocks by computing the cumulative moment density released by aftershocks, considering those located within $l_a=5$km of the fault trace (see the Supplementary Information for the same analysis considering $l_a=2.5$km, Figure S6). We assume that each aftershock, having a seismic moment $M_0$, is a circular crack and compute the slip distribution $\delta(r)$ using \cite{Madariaga13}: 
\begin{eqnarray}
\delta(r) &=& \dfrac{24\Delta \tau}{7\pi\mu} \sqrt{\left(\dfrac{7M_0}{16\Delta \tau}\right)^{2/3}-r^2}
\end{eqnarray}
where $r$ is the crack radius, $\mu = 30$GPa is the shear modulus, and $\Delta \tau$ is the stress drop, assumed to be 3MPa. We bin the aftershock crack radius along-strike, based on the minimum aftershock magnitude for each case (Izmit $M_w$=2.4, Denali $M_w$=1.5, and Craig $M_w$=1.7). Thus the binning sizes employed to discretize the main fault are 90m for Izmit, 32m for Denali, and 40m for Craig earthquake. With the slip distribution, the cumulative seismic moment density of the $i^\textrm{th}$ bin containing $N$ aftershocks is given by
\begin{equation}
m^{(i)}_c  \approx \mu \sum_{j=1}^{N} \delta^{(i)}_{j}
\end{equation}
Note that $\delta^{(i)}_{j}$ is part of the slip distribution of the $j^\textrm{th}$ aftershock projected on the $i^\textrm{th}$ fault segment. We also note that we do not compute the full moment density tensor as we do not have focal mechanisms for all the aftershocks. As we are interested in density relative spatial variation (along the strike of the main fault) of this quantity, the above approximation is adequate. We compute the seismic moment density over two periods following the mainshock: 1 and 3 weeks (Figures \ref{fig2b}a, c, and e). Focusing on the region where the rupture is expected to transition to the supershear regime (based on kinematic models), we systematically observe a small area characterized by a reduced seismic moment density and lack of aftershocks (Figures \ref{fig2b}b, d, and f, pink boxes). The extent of these regions is different for each earthquake c. f., 3.6km for Izmit, 6km for Denali, and 3.2km for Craig. The difference in the smoothness of the cumulative moment curves for different earthquakes is due to the number of aftershocks available for the calculation. In the Craig earthquake case, we have 68 (83) aftershocks at a distance of 5km from the main fault in 1 week (3-week), 2396 (4265) for Denali, and 641 (1118) for the Izmit earthquake. 

\begin{figure}[ht!]
\begin{center}
\includegraphics[width=0.9\textwidth]{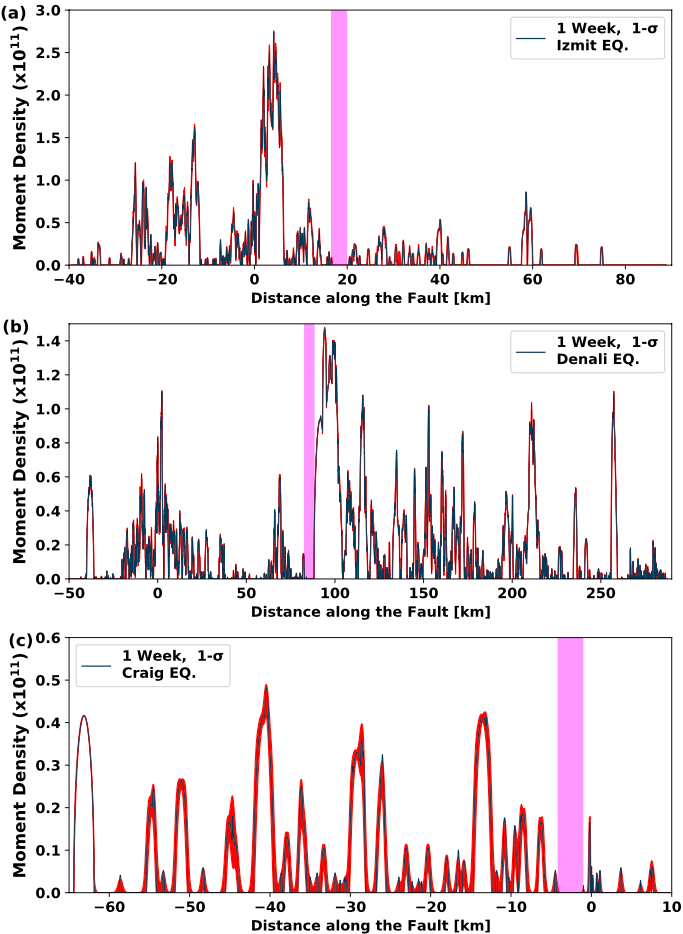}
\end{center}
  \caption{\textbf{High-Resolution Aftershock Catalog Statistical Analysis}. Aftershock seismic moment density projected on the main fault for 1-week, considering all the earthquakes at a distance of 5km from the fault for Izmit (\textbf{a}), Denali (\textbf{b}), and Craig (\textbf{c}) earthquakes. Black curve is the mean seismic moment density on the main fault  $\mathbf{\pm}$ the standard deviation (1-$\mathbf{\sigma}$) computed from 10000 synthetic random catalogs. Pink boxes highlight our proposed transition zone.}
  \label{fig2c}
\end{figure}

In order to evaluate the robustness of our findings, we perform an uncertainty analysis of the seismic moment density estimation by accounting for errors in the location of aftershocks in each catalog. We employ a multivariate normal distribution $\{\mathbf{{x}}\} = \mathcal{N}_{2}(\mathbf{\mu_a},\,\mathbf{\sigma_a})$, assuming that the location uncertainty of aftershocks is normally distributed. Here, $\mathbf{\mu_a}$ corresponds to the aftershock epicenter, while $\mathbf{\sigma_a}$ is the covariance matrix of the location. This procedure allows us to generate, randomly, 10000 synthetic catalogs and repeat the along-strike moment density evaluation with time for each of the 10000 catalogs, deriving the mean and the standard deviation (1-$\sigma$ and 10-$\sigma$) of the spatio-temporal evolution of seismic moment density (red areas in Figure \ref{fig2c} for 1-week, denoting the mean of the seismic moment density on the main fault $\pm$ the standard deviation). Such analysis suggests that despite potential changes in the spatio-temporal distribution of seismic moment density, our conclusions are robust (see the Supplementary Information for a comparison between 1 and 3-week using 1-$\sigma$ and 10-$\sigma$, Figures S7, S8, and S9). 
 
We consider 3-week as the maximum reasonable period to investigate the evolution of the seismic moment for these earthquakes to avoid potential effects of postseismic deformation. Afterslip is proposed as a mechanism driving the aftershock triggering (e.g., Hsu \textit{et al.} \cite{Hsu06} and Ross \textit{et al.} \cite{Ross17}). For each studied event, the inferred transitional region co-locates with areas where some authors report afterslip occurrence \cite{Freed06, Hearn09, Ding15}. Therefore, the observed gap in the early aftershocks productivity (less than 3-week after the mainshock) is mainly related to the mainshock rupture. It is worth noting that results for the $M_{w}$~7.8 Kunlun earthquake reported by Robinson \textit{et al.} \cite{Robinson06}, employing Harvard CMT solutions, also allude to the same conclusion. However, due to the lack of high spatio-temporal density of aftershocks in their catalog, we are more confident in our results using optical correlation techniques to characterize the transition zone. 

We note that in the distribution of moment plotted for the entire fault profile (Figure \ref{fig2b}), local gaps in aftershock density exist and do not necessarily indicate a supershear transition. We hypothesize that such gaps could also be related to abrupt local changes in rupture speeds, off-fault medium strength and fault geometrical complexities which warrants further investigation. 

To conclude, we show that the reduction in the spatial extent of early aftershocks ($\le$ 3-week after the mainshock), and the cumulative seismic moment density around the fault ($\le$ 5km on either side of the fault) is associated with the transition to supershear speed. This type of analysis also helps us identify, more precisely than kinematic models, the region where supershear transition occurs.

\section{Conclusion and Discussion}

Using theoretical arguments and numerical models, that account for coseismic off-fault damage, we have shown that supershear transition is characterized by a significant reduction in the width of the off-fault damage zone. This feature is due to a Lorentz-like contraction of the spatial extent of the stress field around a rupture tip. We then cross-validated this phenomenon with natural observations of coseismic off-fault damage zone width using image correlation and analysis of aftershock catalogs. We confirm that supershear transition is indeed characterized by a significant reduction in the width of the off-fault damage zone. 

Our results are in general agreement with the published kinematic models for the Izmit \cite{Bouchon01, Bouchon02}, Denali \cite{Ellsworth04, Dunham04}, Craig \cite{Yue13}, and Kunlun \cite{Bouchon03_2, Robinson06, Vallee08} earthquakes, relative to the location where the rupture accelerates to supershear speeds. Our inferred location of the transition zone can sometimes be different from those found in kinematic inversions but only by a few kilometers. 

While the theoretical and numerical models are still idealized (modeled on a 1D planar fault with uniform traction and frictional strength), the reduction in the width of the coseismic off-fault damage zone, related to supershear transition, is observed for natural earthquakes (Figures \ref{fig3} and \ref{fig2b}). This approach could thus be used to refine and precisely define the zone of supershear transition inferred from coarser kinematic models. Moreover, as the zone of supershear transition is now narrowed down to a few kilometers, it could also be used to guide future fieldwork to study the segment of the fault where such a transition occurred. A better understanding of the physical conditions for supershear transition might help us foresee the location of future supershear transition, although, as of now, this remains a difficult task.

The lack of aftershocks on the segment experiencing supershear rupture has been debated in the past. Bouchon \& Karabulut \cite{Bouchon08}, for instance, already concluded that the entire supershear segment collocated with a region of aftershock quiescence. To explain this, they claimed that the stress and strength conditions were more homogeneous on the supershear segments \cite{Bouchon08}, impeding large ground accelerations near the fault \cite{Bouchon01, Ellsworth04}. Our work, in contrast, focuses on characterizing the region where the rupture transitions to supershear speed, before the said supershear segment.

The results of this study are valid for a well-developed sub-Rayleigh rupture that transitions to supershear speeds. However, the 2018 $M_w$ 7.5 Palu (Indonesia) earthquake might have either nucleated directly at a supershear speed, or transitioned very early \cite{Bao19, Socquet19, Zhang19, Li20}. The model we develop here might provide insights to understand whether an earthquake can nucleate and propagate directly at supershear speeds, and if so, why that would be the case.

\section*{Data Access}{All the catalogs and numerical simulation results employed in this work have been obtained from published works, cited in the main text and Supplementary Information.}

\section*{Author Contributions}{'HSB conceived of and designed the study. HSB developed the LEFM solution. MYT conducted damage mechanics based numerical modeling and KO performed FDEM based numerical modeling. SA conducted the image correlation analysis. JJ analysed the aftershock catalog with inputs from MYT and HSB. JJ and LB drafted the first version of this manuscript that was subsequently modified with inputs from HSB, MYT, KO, SA, YK, AJR and RJ. All authors read and approved the manuscript'.}

\section*{Funding}{HSB acknowledges the European Research Council grant PERSISMO (grant 865411) for partial support of this work. JJ and RJ acknowledge the funding of the European Research Council (ERC) under the European Union's Horizon 2020 research and innovation program (grant agreement 758210, project Geo4D). LB thanks the funding from the People Programme (Marie Curie Actions) of the European Union's Seventh Framework Programme (FP7/2007-2013) under REA grant agreement PCOFUND-GA-GA-2013-609102, through the PRESTIGE program coordinated by Campus France. SA and YK are partly supported by the ANR project DISRUPT (ANR-18-CE31-0012). R.J. acknowledges funding from the Institut Universitaire de France. Portions of this research were obtained using resources provided by the Los Alamos National Laboratory Institutional Computing Program, which is supported by the U.S. Department of Energy National Nuclear Security Administration under Contract No. 89233218CNA000001. This publication was approved for unlimited release under LA-UR-21-24016. SPOT images are from the ISIS program from CNES.}

\clearpage
\begin{center}
\Large{\noindent \textbf{SUPPLEMENTARY INFORMATION}}\\[12pt]
\end{center}
\renewcommand\thesection{S\arabic{section}.~}
\renewcommand\thefigure{S\arabic{figure}}
\renewcommand\thetable{S\arabic{table}}
\setcounter{figure}{0}
\setcounter{section}{0}
\setcounter{table}{0}

\section{Numerical modeling of off-fault coseismic damage}

Specific parameters used in the FDEM model are provided in table~\ref{tab:FDEM_param} and those used in the micromechanics based model are provided in table~\ref{tab:micro_param}.

\begin{table}[ht!]
\caption{Parameters for contact interactions within the off-fault medium, FDEM model.}
\begin{center}
\begin{tabular}{c l c}
\hline
Parameters & Description & Values \\
\hline
$G_{IC}$ & Fracture energy for tensile cohesion (kJ/m$^2$) & 3.0\\
$C_{I}$ & Cohesion for mode I opening crack (MPa) & 8.0\\
$\delta_{c,I}$ & Critical cohesive weakening displacement in tension (mm)  & 0.75 \\
$G_{IIC}$ & Fracture energy for shear cohesion (kJ/m$^2$) &20.0\\
$C_{II}$ & Cohesion for mode II shear crack (MPa) & 27.5\\
$\delta_{c,II}$ & Critical cohesive weakening displacement in shear (mm) & 1.5 \\
$f_{s,o}$ & Static friction coefficient & 0.6 \\
$f_{d,o}$ & Dynamic friction coefficient & 0.1 \\
$G_{IIC}^f$ & Shear fracture energy for friction (kJ/m$^2$)  & 20.0\\
$\delta_{II}^{f}$ & Critical weakening displacement for friction (mm) & 1.67
\end{tabular}
\label{tab:FDEM_param}
\end{center}
\end{table}

\begin{table}[ht!]
\caption{Parameters used for the micromechanical model}
\begin{center}
\begin{tabular}{c l c}
\hline
Parameters & Description & Values \\
\hline
$a_0$ & Penny shape crack radius (m) & $60$\\
$N_v$ & Volume density of cracks ($\times10^{-7}$ \#/m$^3$)&  $1.68$  \\
$D_0$ & Initial damage density &  $0.1$  \\
$v_m$  & branching speed	(km/s) &	$1.58$ \\
$\beta$ & Ashby \& Sammis \cite{Ashby90} factor	& 0.1 \\
$\Omega$ & Crack factor	&	  2.0 \\
$t^*$ &  Prakash \& Clifton \cite{Prakash93} time (s) &$40\times10^{-3}$ \\
\end{tabular}
\label{tab:micro_param}
\end{center}
\end{table}

\begin{figure}[ht!]
\begin{center}
\includegraphics[width=0.9\textwidth]{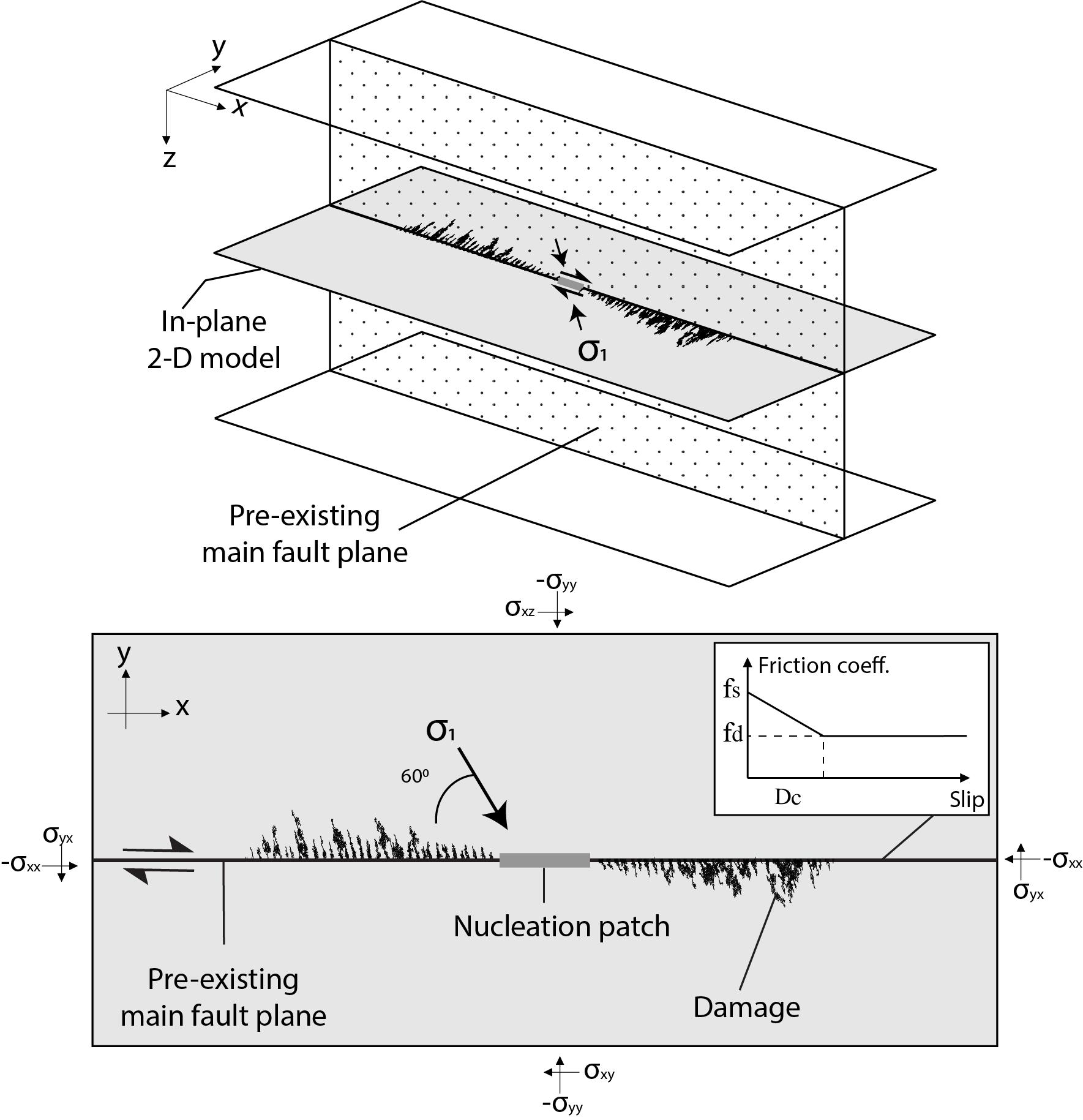}
\end{center}
\caption{Schematics and parameters used for the simulations of dynamic ruptures in a 2-D in-plane model. We consider a right-lateral planar fault,  embedded in a brittle off-fault medium (FDEM or micromechanical model).  Slip-weakening friction (grey box) acts along the main fault plane.The medium is loaded by uniform background stresses with the maximum compressive stress $\sigma_1$ making an angle of $60^o$ with the fault plane. The thick grey line corresponds to the nucleation zone where either the initial shear stress is set-up to be just above the fault strength (micromechanical model), or we apply a local decrease of the static friction (FDEM model). Figures adapted from \cite{Thomas17_2} and \cite{Okubo19}}.
\end{figure}

\begin{figure}[ht!]
\begin{center}
\includegraphics[width=1\textwidth]{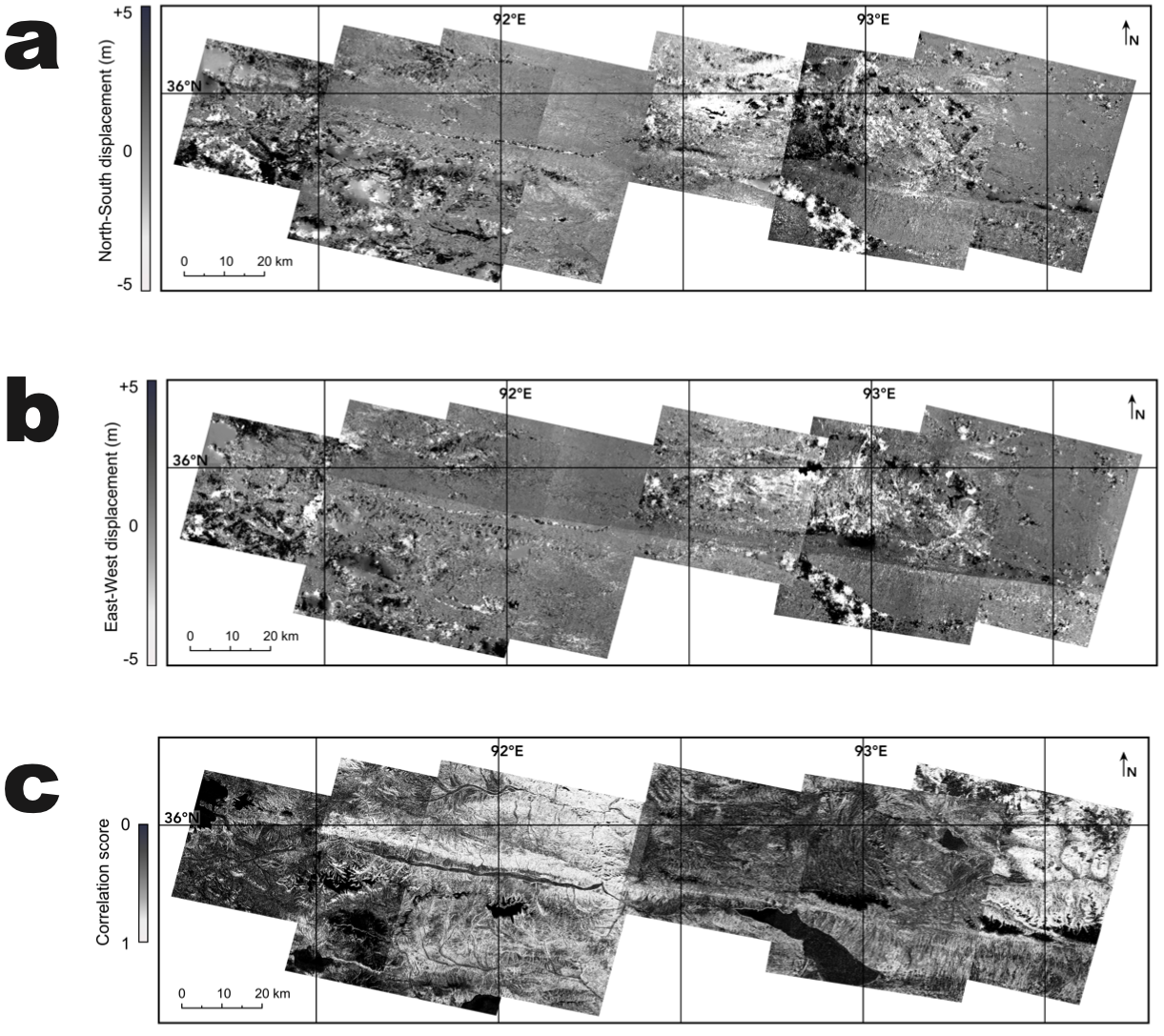}
\end{center}
\clearpage
\caption{(a) North-South and (b) East-West surface displacement maps for the strike-slip part of the 2001 Kunlun surface rupture. Results benefit from a 10-meters ground resolution after the horizontal correlation of SPOT (1 to 4) pre- and post-earthquake images using MicMac. North-South displacements are close to zero at the fault while there is  a clear left-lateral offset in the East-West displacement map. (c) Correlation score provided by MicMac after the horizontal correlation of the SPOT images. Dark areas correspond to decorrelation in lakes, drainages or snow. White areas, representing areas of good correlation, are present around the fault zone and particularly between 91.5 and 92.5$^{o}$ E.}
\end{figure}

\begin{figure}[ht!]
\begin{center}
\includegraphics[width=1.0\textwidth]{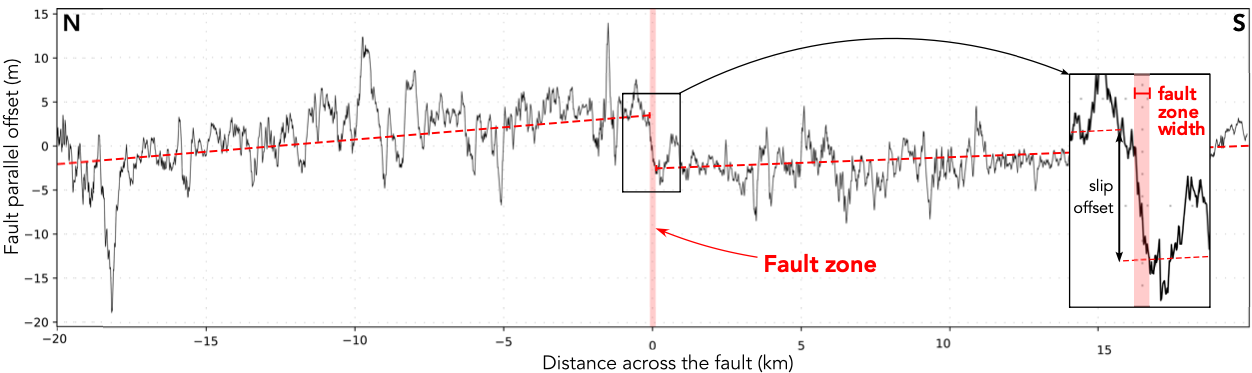}
\end{center}
\clearpage
\caption{Example of one of the 396 profiles obtained from the optical correlation image processing along the fault. The step in the figure represents the coseismic offset produced by the earthquake, while the red box denotes the region defined during this work as the width of the off-fault damage zone.}
\end{figure}

\begin{figure}[ht!]
\begin{center}
\includegraphics[width=0.85\textwidth]{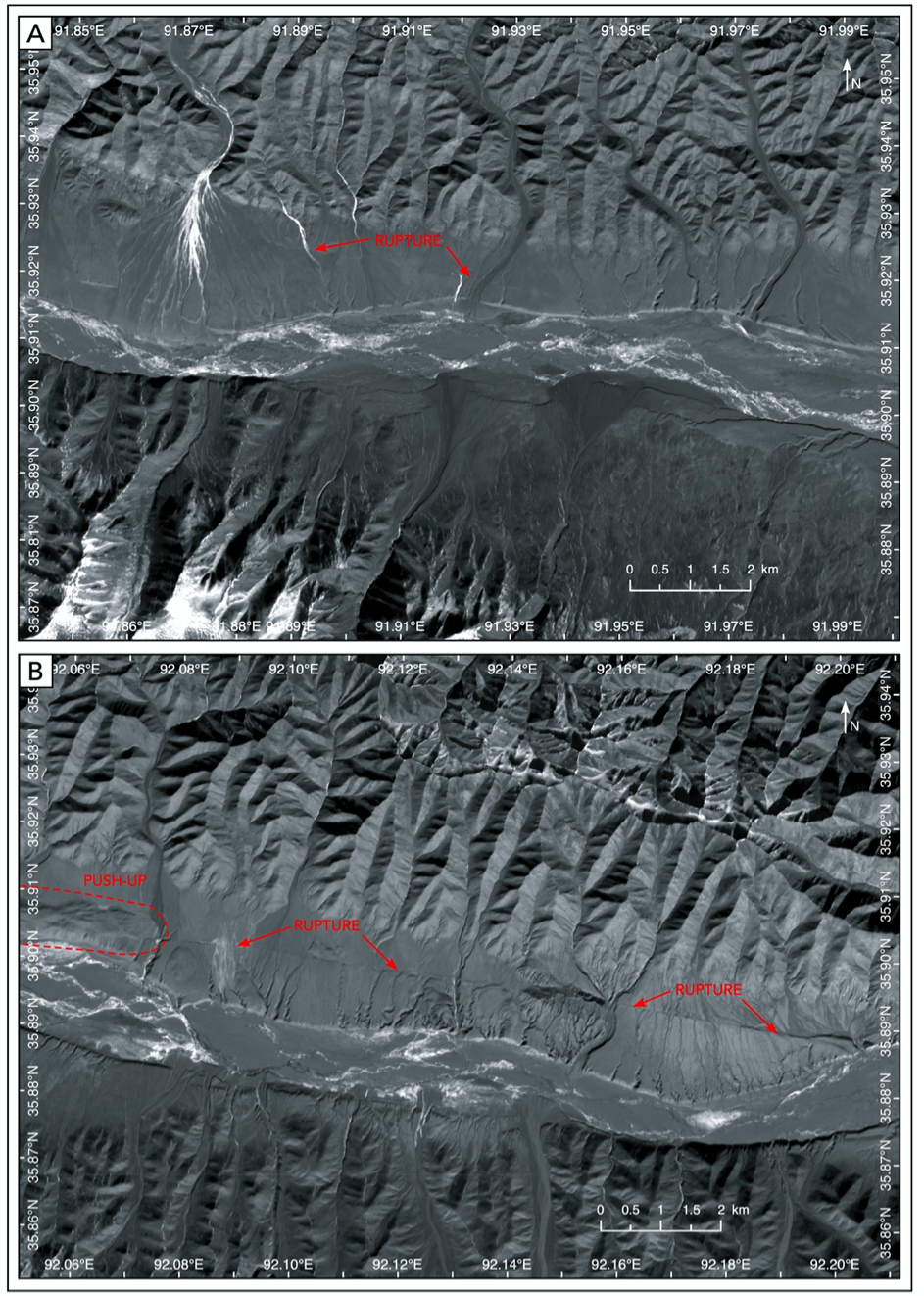}
\end{center}
\clearpage
\caption{SPOT image of the 2001 Kunlun rupture area: (A) before the rupture transitioned to supershear speed, and (B) after the rupture transitioned to supershear speed. The rupture is indicated in red at some locations. The rupture trace is located in some places with red arrows. The rupture is clean in image B), unlike in A) where the trace is more ambiguous.}
\end{figure}

\begin{figure}[ht!]
\begin{center}
\includegraphics[width=1.\textwidth]{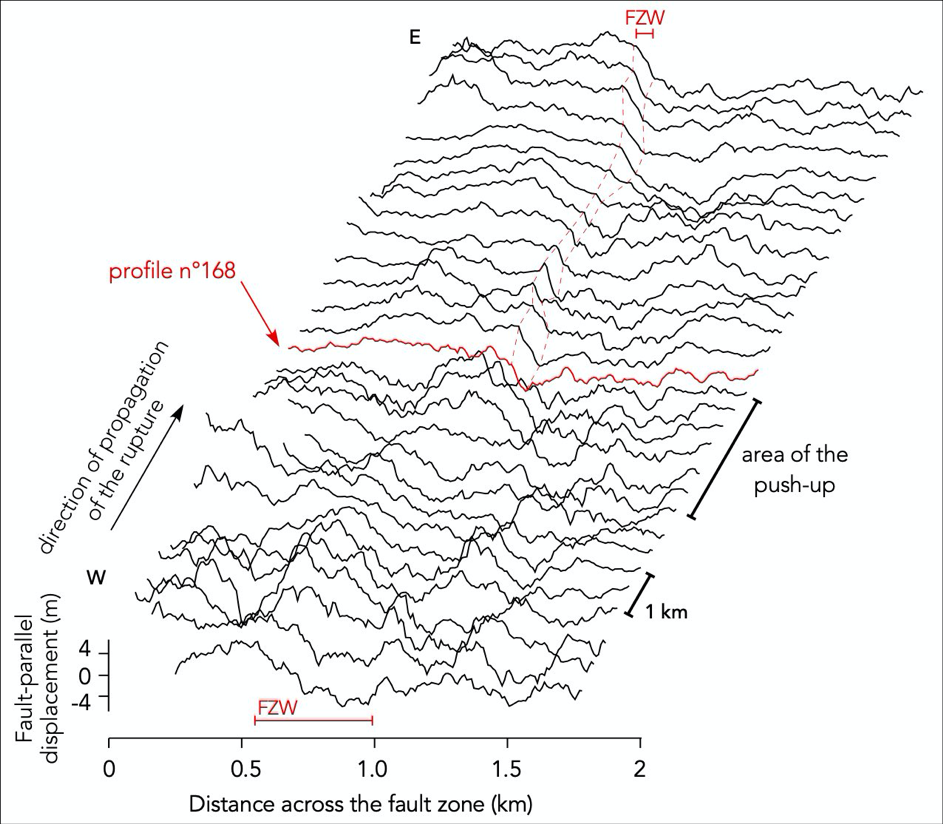}
\end{center}
\clearpage
\caption{Across-fault displacement profiles showing the fault-parallel component of displacement (left-lateral displacement). Profiles are arranged in a three-dimensional view. Plot axes show displacement amplitude and distance across the fault for the first profile to the bottom of the plot. The rupture propagates from the west to the east, from the bottom to the top of the plot. Profile 168 corresponds to the profile where we observe a localized fault zone relative to the westward area.  Eastward to profile 168, the localized fault zone width is indicated by light-red dashed lines. Westward to profile 168, the fault zone width is not well defined, as seen in the optical images (IMAGE B, Figure S4).}
\end{figure}

\begin{figure}[ht!]
\begin{center}
\includegraphics[width=0.75\textwidth]{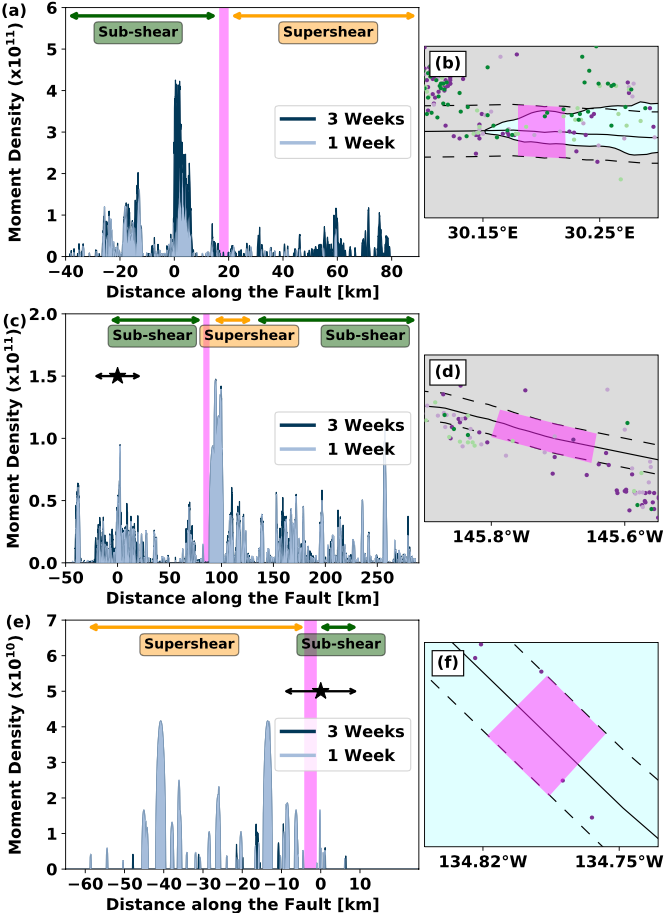}
\end{center}
  \caption{\textbf{Spatio-Temporal Seismic Moment Density Evolution}. Cumulative aftershock seismic moment density projected on the main fault at different temporal scales (1-3 weeks), for Izmit (\textbf{a}), Denali (\textbf{c}), and Craig (\textbf{e}) earthquakes. All the aftershocks within a distance of 2.5~km from the fault are considered in the calculation (area denoted by the black discontinuous lines in Figures \textbf{6a}, \textbf{b} and \textbf{c} on the main text). Color-coded arrows (on top of \textbf{a}, \textbf{b}, and \textbf{c}) indicate the different speed regimes reported for each event (green for sub-Rayleigh and orange for supershear) \cite{Bouchon08,Ellsworth04,Yue13}, while the starts denote the epicenter of each earthquake and the arrows indicate the ruptures' direction. The pink boxes point out our proposed transition zone, also observed in a map view in \textbf{b} for Izmit, \textbf{d} for Denali, and \textbf{f} for Craig earthquakes.}
\end{figure}

\begin{figure}[ht!]
\begin{center}
\includegraphics[width=0.6\textwidth]{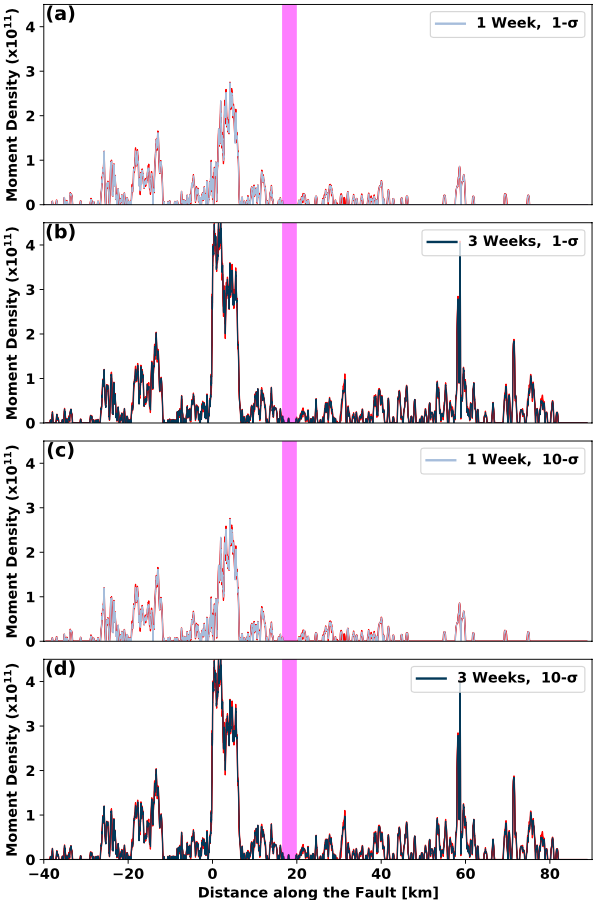}
\end{center}
\clearpage
\caption{\textbf{High-Resolution Aftershock Catalog Statistical Analysis, Izmit Earthquake, 1-and 10-$\mathbf{\sigma}$}. \textbf{a} and \text{b} show the aftershock seismic moment density projected on the main fault at different temporal scales (1-3 weeks), considering all the aftershocks at a distance of 5 km from the fault. The red area in each panel denotes the mean seismic moment density projected on the main fault $\mathbf{\pm}$ the standard deviation (1-$\mathbf{\sigma}$, indicated in the legend) of the 10000 synthetics catalogs performed for the analysis. The pink box indicates this work's proposed transition zone. \textbf{c} and \textbf{d} are the same plots than before (\textbf{a} and \textbf{b}), but considering a 10-$\mathbf{\sigma}$ standard deviation on the calculation. }
\end{figure}

\begin{figure}[ht!]
\begin{center}
\includegraphics[width=0.6\textwidth]{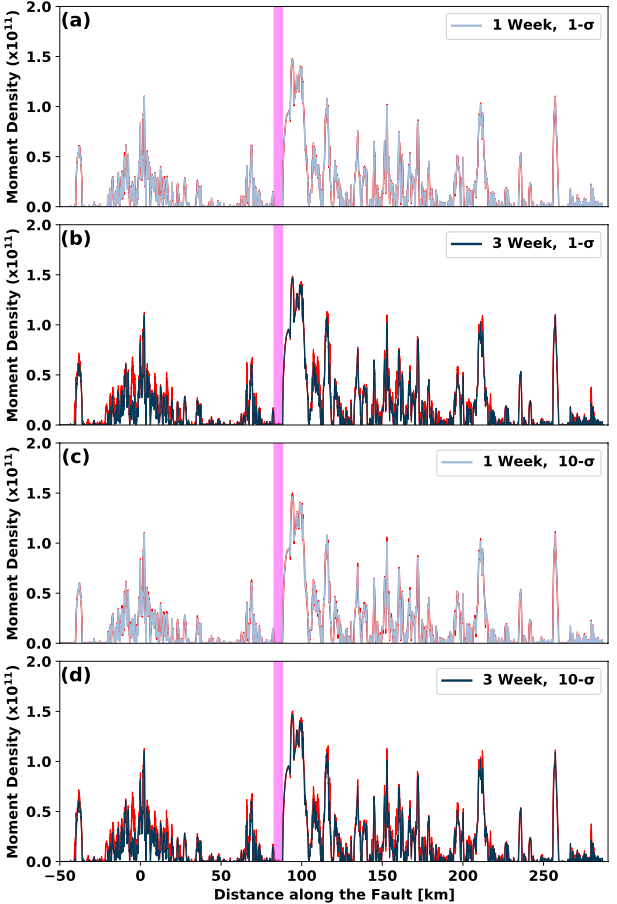}
\end{center}
\clearpage
\caption{\textbf{High-Resolution Aftershock Catalog Statistical Analysis, Denali Earthquake, 1-and 10-$\mathbf{\sigma}$}. \textbf{a} and \text{b} show the aftershock seismic moment density projected on the main fault at different temporal scales (1-3 weeks), considering all the aftershocks at a distance of 5 km from the fault. The red area in each panel denotes the mean seismic moment density projected on the main fault $\mathbf{\pm}$ the standard deviation (1-$\mathbf{\sigma}$, indicated in the legend) of the 10000 synthetics catalogs performed for the analysis. The pink box indicates this work's proposed transition zone. \textbf{c} and \textbf{d} are the same plots than before (\textbf{a} and \textbf{b}), but considering a 10-$\mathbf{\sigma}$ standard deviation on the calculation. }
\end{figure}

\begin{figure}[ht!]
\begin{center}
\includegraphics[width=0.6\textwidth]{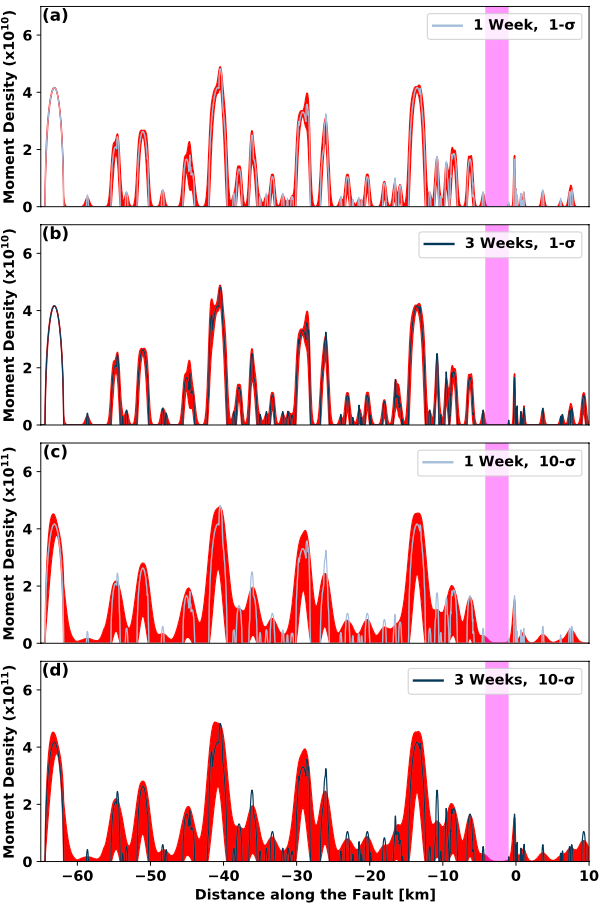}
\end{center}
\clearpage
\caption{\textbf{High-Resolution Aftershock Catalog Statistical Analysis, Craig Earthquake, 1-and 10-$\mathbf{\sigma}$}. \textbf{a} and \text{b} show the aftershock seismic moment density projected on the main fault at different temporal scales (1-3 weeks), considering all the aftershocks at a distance of 5 km from the fault. The red area in each panel denotes the mean seismic moment density projected on the main fault $\mathbf{\pm}$ the standard deviation (1-$\mathbf{\sigma}$, indicated in the legend) of the 10000 synthetics catalogs performed for the analysis. The pink box indicates this work's proposed transition zone. \textbf{c} and \textbf{d} are the same plots than before (\textbf{a} and \textbf{b}), but considering a 10-$\mathbf{\sigma}$ standard deviation on the calculation. }
\end{figure}

\clearpage
\bibliography{references}
\bibliographystyle{naturemag}

\end{document}